\documentclass[usenatbib]{mn2e}
\usepackage{txfonts,graphicx}
\usepackage{amssymb}
\newcommand{\hi}{H\,{\sc i}}
\newcommand{\hii}{H\,{\sc ii}}
\newcommand{\ha}{\ensuremath{\mbox{H}\alpha}}
\newcommand{\hb}{\ensuremath{\mbox{H}\beta}}

\newcommand{\heii}{He\,{\sc ii}}
\newcommand{\oi}{O\,{\sc i}}
\newcommand{\oii}{O\,{\sc ii}}
\newcommand{\oiii}{O\,{\sc iii}}

\newcommand{\sii}{S\,{\sc ii}}
\newcommand{\nii}{N\,{\sc ii}}
\newcommand{\neiii}{Ne\,{\sc iii}}
\newcommand{\nev}{Ne\,{\sc v}}
\newcommand\aj{{AJ}}% 
\newcommand\araa{{ARA\&A}}% 
\newcommand\apj{{ApJ}}% 
\newcommand\apjl{{ApJ}}% 
\newcommand\apjs{{ApJS}}% 
\newcommand\aap{{A\&A}}% 
\newcommand\mnras{{MNRAS}}% 
\newcommand\nat{{Nature}}% 
\catcode`\@=11
\def\gapprox{\mathrel{\mathpalette\@versim>}}
\def\lapprox{\mathrel{\mathpalette\@versim<}}
\def\@versim#1#2{\lower2.45pt\vbox{\baselineskip0pt\lineskip0.9pt
      \ialign{$\m@th#1\hfil##\hfil$\crcr#2\crcr\sim\crcr}}}
\catcode`\@=12
\newcommand\phn{\phantom{0}}% 
\newcommand\phs{\phantom{$-$}}% 
\newcommand{\flux}{\ensuremath{\,\mbox{erg}\,\mbox{cm}^{-2}\mbox{s}^{-1}}}
\newcommand{\pccm}{\ensuremath{\,\mbox{cm}^{-3}}}

\title{Emission-line Diagnostics of Low Metallicity AGN}
\author[B.~Groves et al.]{Brent A.~Groves$^{1}$, Timothy M.~Heckman$^2$ and Guinevere
Kauffmann$^1$\\
$^1$Max-Planck Institut f\"ur Astrophysik, Karl-Schwarzschild
Strasse 1, Garching D-85748, Germany\\
$^2$Dept. of Physics and Astronomy, John Hopkins University,
Baltimore, MD 21218, USA}

\begin{document}

\maketitle

\begin{abstract}
Current emission-line based estimates of the metallicity of active
galactic nuclei (AGN) at both high and low redshifts indicate that AGN
have predominantly solar to supersolar metallicities. This leads to
the question: do low metallicity AGN exist? In this paper 
we use photoionization models
to examine the effects of  metallicity variations on the
narrow emission lines from an AGN. We explore a variety of
emission-line diagnostics that are  useful for identifying
AGN with low metallicity gas. 
We find that line ratios involving [\nii]
are the most robust metallicity indicators in galaxies where the primary source
of ionization is from the active nucleus. Ratios involving
[\sii] and [\oi] are strongly affected by uncertainties in
modelling the density structure of the narrow line clouds.
To test our diagnostics, we  turn to an analysis of AGN in
the Sloan Digital Sky Survey (SDSS).
We find a clear trend in the relative strength of [\nii] with
the mass of the AGN host galaxy. The metallicity of the ISM
is known to be correlated with stellar mass in star-forming
galaxies; our results indicate that a similar trend exists for
AGN. We also find that the best-fit models for  typical Seyfert
narrow line regions have  supersolar abundances.
Although there is  a mass-dependent range of a factor of 2-3 in the 
NLR metallicities of the AGN in our sample, 
AGN with sub-solar metallicities are very rare in the SDSS.
Out of a sample of $\sim$23000 Seyfert 2 galaxies
we find only $\sim$40 clear candidates for  AGN with NLR
abundances that are below solar. 
\end{abstract}


%\doublespace
\section{Introduction}

In recent years there has been increasing  evidence that the
growth of supermassive black holes at the centres of galaxies 
is closely linked to nuclear star
formation and the formation of spheroids
\citep[eg][]{Richstone98,Haehnelt98}. The tight correlation between the
black hole mass and bulge velocity dispersion
\citep{Ferrarese00,Gebhart00}, as well as the weaker correlation
between black hole mass and spheroid mass
\citep{Kormendy95,Magorrian98}, indicate that the fuelling of the
central black hole must be accompanied by spheroid growth. Although
part of the
spheroid growth may result  from mergers of stellar disks 
\citep[eg][]{Kauffmann00},  star
formation in the central regions of the galaxy  is also expected to 
contribute to the overall growth of the bulge
 \citep{Schmitt99,CidFernandes01,CidFernandes04}.

Active galactic nuclei (AGN) are found in galaxies where black holes are
growing through the accretion of surrounding gas. 
The gas is also likely to be associated with the nuclear star formation.
By measuring the element abundances of the gas surrounding the AGN we
obtain an indirect measurement of the star formation history of the
host galaxy. If most of the gas has been converted into stars before
the AGN becomes visible, then the gas is likely 
to be of high metallicity. If,
on the other hand, the AGN is concurrent with or precedes the main
episode of star formation in the bulge, low metallicity   
gas is likely to be found in some AGN.

There have been analyses of elemental abundances in AGN for over
two decades \citep{Davidson79}. In high-$z$ AGN, this work has mainly
focused on  the broad emission lines to determine abundances. There
have also been a number of studies that have 
used absorption line and narrow emission line diagnostics
\citep[see][ for a review]{Hamann99}. In nearby AGN the narrow
emission lines are readily observable, and 
are often the easiest method to determine
metallicities as pioneered by \citet{Storchi89}. 
The results for both high and low redshift AGN appear to show that
most AGN have solar to supersolar metallicities
\citep{Storchi98,Hamann02}.  Although the results do depend on   
the detailed model assumptions
\citep[see
eg][]{Hamann02,Komossa97}, there does appear to be some  consensus that
the gas in AGN is usually of  high metallicity. 

The paucity of AGN with sub-solar metallicities  may be linked to the fact
that AGN in low mass galaxies appear to be
very rare in the local Universe. 
\citep[see
e.g.][]{Greene04,Barth05}. Gas-phase metallicity is known
to be strongly correlated with stellar mass \citep{Tremonti04}.
If AGN are predominantly found in massive, bulge-dominated
galaxies that have processed most of their gas
into stars by the present epoch, it may not be too
surprising that the inferred metallicities are almost always high.  

Although these conditions may hold at the present epoch, it is
interesting to speculate whether
the host galaxies of  AGN might  be considerably 
different at higher redshifts. For example,
in the models of \citet{Kauffmann00} the typical gas fractions
in AGN hosts are expected to evolve from $5-10 \%$ at the present
day, to $\sim 20\%$ at $z=1$ and $\sim 50\%$ at $z\sim 2$.
Even if high redshift  AGN reside in massive galaxies,
the gas-phase metallicities of their hosts  might still be expected 
to be lower than at the present day.

In this paper, we explore these questions by using photoionization models
to examine the effects that a  variation in metallicity would have upon the
narrow emission lines from an AGN. We ask whether there are
particular emission-line  diagnostics that are most useful for identifying
AGN with low metallicity gas.
Using the  models as a guide, we then search for low
metallicity AGN in the Sloan Digital Sky
Survey (SDSS). The SDSS provides high quality  spectra from
which we can measure both the host galaxy and the emission-line properties   
of a very large sample of nearby AGN, 
so it is an ideal database for searching for rare cases of low
metallicity AGN in the nearby Universe. Finally, we discuss the
implications of our results for searches for low metallicity
AGN at higher redshifts.

\section{Narrow Line Regions and Low Metallicity}

The use of nebular emission lines produced in \hii\ regions to
infer the gas-phase abundances of extragalactic objects has a long history 
\citep[eg][]{Aller42}. Once understanding grew of the nuclear
emission from AGN it was 
not long before the use of emission lines as abundance diagnostics was
extended to these objects
\citep[eg][]{Davidson79,Storchi89}. 

There is a good understanding of the 
ionizing sources in \hii\ regions (the O \& B stars) and 
photoionization models to treat the physics of these
regions are quite well-developed. As a result,  
the correlation of emission line ratios with
abundances in star-forming regions is reasonably well
understood. Recent papers by 
\citet{Charlot01} and \citet{Kewley02} have provided
calibrated relationships between gas-phase oxygen abundance 
and combinations of strong   emission
lines that are observable at optical wavelengths.

For AGN the situation is markedly different. The ionizing source is not
as well understood, and has a much greater range in luminosity.
The ionizing spectrum is
also harder than that produced by stars, and shocks are also more likely to
contribute to the ionization state. While recent theoretical models
\citep{Baldwin95,Groves04a} have helped clarify some of the
ionization structure, considerable uncertainties still exist.  
Many AGN show both broad and narrow line emission, and  the
ionization conditions in both the broad and narrow emission line
regions show strong variations between different AGN. 

Both the broad and the narrow emission lines can be used to estimate gas
phase abundances. The broad line region (BLR) samples the
gas closer to the nucleus. 
However, these emission lines are only seen in Type 1 AGN , and
the strong emission lines used for abundance analysis are only found
in the UV. As a result, abundance analysis is restricted to high redshift AGN
or to objects with UV observations from space \citep[see e.g.][]{Hamann99}.

The narrow line regions (NLR) of AGN are more readily observable. Their
greater spatial extent means they are less likely to be obscured. 
The NLR also produces  strong optical emission lines from
several species. The physics of the NLR is also  
simpler than that of the BLR; the physical conditions are in fact
similar to \hii\ regions, so that many of the                 
metallicity-sensitive emission line ratios that are routinely applied
to star forming galaxies are also good diagnostics in the NLR.
\citet{Storchi98} have  calibrated several of these
emission line ratios for nearby AGN, using \hii\ region determined
metallicities and photoionization models.  
The similarity between the NLR and the  \hii\ regions in a galaxy   
also leads to ambiguities when estimating abundances. Unless
spatially-resolved observations can be obtained, it is often
difficult to disentangle the contributions to the emission line
spectrum from the NLR and from star-forming regions within the galaxy.

\begin{figure}
\includegraphics[width=\hsize]{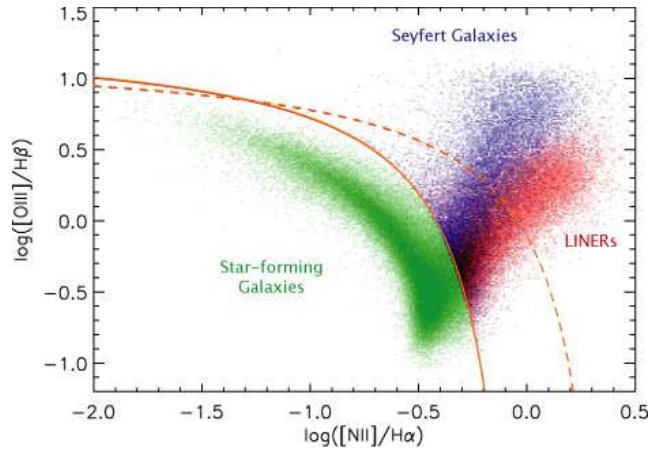}
\caption{The \citet{BPT81} emission line ratio diagnostic diagram of
[\nii]$\lambda 6584$\AA/\ha\ versus [\oiii]$\lambda 5007$\AA/\hb\
is plotted for SDSS emission line galaxies.
Star-forming galaxies are shown in  green, Seyfert 2s
in blue and LINERs in red. The two different curves
used to separate AGN and star bursts are indicated by the  
solid curve (Kauffmann separator) and dashed
curve (Kewley separator).
}\label{fig:BPTeg}
\end{figure}

Figure \ref{fig:BPTeg} shows one of the standard line ratio diagrams
\citep{BPT81} 
used to distinguish  
star-forming galaxies, Seyfert 2 galaxies (or AGN NLRs) and Low
Ionization Nuclear
Emission-line Region galaxies (LINERs). A detailed description
of the location of these different classes of object in the
different line ratio diagrams and the criteria that
define our classification system has been given in a recent
paper by \citet{Kewley06}. 

In this particular diagram, the emission line galaxies are distributed in
$V$-shaped morphology, with the star-forming galaxies located in the left
branch and the AGN in the right
branch. The AGN branch actually consists of two classes of object, the
Seyfert 2 galaxies and the LINERs \citep{Kewley06}. 

The distribution of the starforming galaxies along the
horizontal [\nii]$\lambda 6584$\AA/\ha\ axis is strongly correlated
with the metallicity within the starforming region and the mass of the
galaxy; both increase with increasing [\nii]/\ha. The vertical axis
of [\oiii]$\lambda 5007$\AA/\hb\ is associated with the average
ionization state and temperature of the photoionized gas in the
emission line galaxy. AGN are offset from the
star-forming galaxies on this axis because of their much harder ionizing
spectrum.
 Those objects which lie at the
extremities of the AGN branch are ``pure'' AGN. As the contribution of
star-formation to the emission line spectrum increases, the strength
of [\oiii]/\hb\ decreases. This AGN-starburst
mixing sequence culminates  at the fulcrum of the $V$,
where starforming galaxies and AGN dominated by starformation  become
indistinguishable. 
When we model NLR  spectra, we consider only ``pure'' AGN.

\section{Low Metallicity NLR simulations}

We have run a series of narrow line region (NLR) models with
a wide range in both metallicity and
ionization parameter\footnote{The ionization parameter is a measure of
the density of ionizing photons relative to the gas density.}.
The models are based upon the dusty, radiation pressure dominated models of
\citet{Groves04a,Groves04b}, but use a more realistic ionizing spectrum
\citep{Groves06}. While it is unknown whether low metallicity AGN will
contain dust, the observations of dust in high-$z$ objects indicates
that such an assumption is not unreasonable \citep{Bertoldi03}.
Unlike previous work on NLR metallicity \citep[eg][]{Storchi98}, these
models assume a more 
physical structure for the NLR clouds and concentrate upon the effects
of low metallicity.

As the models have been discussed in depth in previous papers, we
only describe the relevant parameters and the changes that were made 
in order to address the issue of low metallicity AGN.
Throughout this work we assume a fiducial value of $n_{\mathrm{H}}\sim
10^3$\pccm\ for the density as
found in previous papers.

\subsection{Metallicity and Dust Depletion}

The assumed  abundances  are based upon the 
solar abundance set of \citet{Asplund05}, which takes into account
recent revisions of the solar abundances of several important elements
like carbon and oxygen.
Note that abundance ratios may vary between different AGN  and even within
the NLR of a single AGN; the abundances given in table \ref{tab:elements} 
are thus meant to be representative.

\begin{table}
\begin{center}
\caption{Solar abundance \& metallicity scaling \label{tab:elements}}
\begin{tabular}{lll}
\hline
Element & Abundance\footnotemark &
Depletion\footnotemark\\
\hline
\hline
H .................................. & \phs0.000\phn   & \phs0.00\phn\\
He.................................. & $-$0.987\phn    & \phs0.00\phn\\
C .................................. & $-$3.61\phn\phn & $-$0.30\phn \\
N .................................. & $-$4.20\phn\phn & $-$0.30\phn \\
O .................................. & $-$3.31\phn\phn & $-$0.24\phn \\
Ne.................................. & $-$3.92\phn\phn & \phs0.00\phn\\
Na.................................. & $-$5.68\phn\phn & $-$0.60\phn \\
Mg.................................  & $-$4.42\phn\phn & $-$1.15\phn \\
Al.................................. & $-$5.51\phn\phn & $-$1.44\phn \\
Si...................................& $-$4.49\phn\phn & $-$0.89\phn \\
S .................................. & $-$4.80\phn\phn & $-$0.34\phn\\
Cl.................................. & $-$6.72\phn\phn & $-$0.30\phn \\
Ar.................................  & $-$5.60\phn\phn & \phs0.00\phn\\
Ca.................................  & $-$5.65\phn\phn & $-$2.52\phn \\
Fe.................................. & $-$4.54\phn\phn & $-$1.37\phn \\
Ni.................................. & $-$5.75\phn\phn & $-$1.40\phn \\
\hline
\end{tabular}
\end{center}
\end{table}
\footnotetext[2]{All abundances are logarithmic with
respect to Hydrogen}
\footnotetext[3]{Depletion given as
$\log(X/\mathrm{H})_{\mathrm{gas}}-
\log(X/\mathrm{H})_{\mathrm{ISM}}$}

As discussed in \citet{Groves04a}, the abundance with respect to
hydrogen of most of the
elements listed in table \ref{tab:elements} will scale linearly with
total metallicity. The exceptions are He and N. 
For helium, the yield from stars must be added to the primordial
abundance. In this work, we use the primordial measurements of
\citet{Pagel92}, which gives                                        
\begin{equation}
{\rm He}/{\rm H}=0.0737+0.0293Z/Z_{\odot}.
\end{equation}

As nitrogen possesses both primary and secondary nucleosynthesis
components, the relationship between nitrogen abundance and metallicity differs
from that of other elements. For low metallicity galaxies (log (O/H)
$\lapprox -4.0$) the N/O ratio is approximately constant, as expected for a
``primary'' element whose production is independent of
metallicity. However, for higher metallicity galaxies (log (O/H) $\gapprox
-3.5$) the N/O ratio is found to rise steeply with metallicity. This
suggests that nitrogen becomes dominated
by secondary production from CNO nucleosynthesis. 

We follow \citet{Groves04a} and use a linear combination
of the primary and secondary components of Nitrogen. This relationship
is fitted to the
data from \citet{Mouhcine02} and
\citet{Kennicutt03}, with the requirement of matching the solar
abundance patterns \citep[see figure 2 in][]{Groves04a}. With the new
abundances in table \ref{tab:elements} we obtain  the following
relation,
\begin{equation}\label{eqn:N/H}
\left(\mathrm{N}/\mathrm{H}\right) =\left(\mathrm{O}/\mathrm{H}\right)
\left(10^{-1.6} + 10^{\left(2.37 +
\log_{10}\left[\mathrm{O}/\mathrm{H}\right]\right)}\right)
\end{equation}

Using these prescriptions, we have then explored six NLR
metallicities:
$4Z_{\odot}$, $2Z_{\odot}$, $1Z_{\odot}$, $0.5Z_{\odot}$, $0.25Z_{\odot}$,
$0.1Z_{\odot}$, and $0.05Z_{\odot}$. This range is wide enough to
explore trends in emission line properties  associated
with a decrease in AGN metallicity. In terms of $12+\log($O/H$)$,
these abundances correspond to 9.29, 8.99, 8.69, 8.39, 8.09, 7.69 and 7.39
respectively. Note that with the new \citet{Asplund05} abundances, the
oxygen abundance at solar is approximately 0.2 dex lower than that
given in previous work \citep{Grevesse98}.
As some of the metals present in the NLR clouds are depleted onto
dust, the actual metallicity of the gas component in our models is
approximately half of the total metallicity.

\subsection{Narrow Line Region Dust}

While the total dust to gas ratio is known to decrease with decreasing
metallicity, the actual variation of dust abundance with metallicity
is uncertain. In AGN no direct measurements can be made of
dust depletion so for simplicity,
we assume
a solar depletion pattern for the dust and
one that is constant with metallicity. 
In Galactic regions with different physical
conditions, the dust to gas ratio, and hence the depletion factors, are
observed to vary \citep{Savage96}. However, when regions of differing
metallicities are observed, the depletion factors appear to be
approximately constant \citep{Vladilo02}. Our use of constant
depletion factors for NLR models of differing metallicities is therefore not
unreasonable. 

The depletion factors, given in table
\ref{tab:elements}, are based upon the recent work of \citet{Kimura03} 
who examined the metal absorption along several lines of sight within
the local bubble.  The gas abundances were 
compared with solar abundances to determine the depletion. Some elements
were  not discussed in this work and for these we use previously assumed
values from \citet{Dopita00}. 
For all other dust properties we follow the \citet{Groves06}
description of NLR dust. Note that any variation of the depletion
factors will have a similar effect  on the
model to abundance variations. 

\subsection{Ionizing Spectrum}

We follow \citet{Groves06}
and use an empirical fit to the \citet{Elvis94}
observations, 
\begin{equation}\label{eqn:AGNspec}
\begin{array}{rcl}
f_\nu &=& \nu^{\alpha_\mathrm{EUV}}
\exp\left(-\frac{h\nu}{kT_\mathrm{UV}}\right)
\exp\left(-\frac{kT_\mathrm{BBB}}{h\nu}\right)\\
&+& a\nu^{\alpha_\mathrm{X}}
\exp\left(-\frac{h\nu}{kT_\mathrm{X}}\right)
\exp\left(-\frac{kT_\mathrm{BBB}}{h\nu}\right).
\end{array}
\end{equation}
with  $\alpha_\mathrm{EUV}=-1.75$, 
$kT_\mathrm{UV}=120$ eV, $kT_{\mathrm{BBB}}= 7.0$ eV. 

The X-ray part of the spectrum has an index of
$\alpha_\mathrm{X}=-0.85$ and an upper cut-off of $kT_\mathrm{X}=10^5$ eV.
The parameter $a$ is set to 
$a=0.0055$, which results in an optical-X-ray 
slope of index $\alpha_\mathrm{O-X}\sim-1.4$.  
For simplicity, we use the same ionizing spectrum for AGN 
of all metallicities, but we note that a 
lower abundance of metals and especially dust is likely to alter the
structure of the AGN torus and possibly even the accretion disk.

\subsection{NLR model Parameters}

For each metallicity we model a range of NLR ionization conditions. We
follow the work of \citet{Groves06} and maintain a constant total
pressure across the models, while varying both the incident flux
density and initial gas pressure. The total pressure we consider is
$P_{\rm tot}/k =10^7$K\pccm. This corresponds to a [\sii] density of
$n_{\rm H}\simeq 10^3$\pccm\ which is a reasonable estimate of the
NLR number density. While the effect of radiation pressure upon dust
becomes  less important at low metallicities, we maintain the
same input parameters across the models for comparison purposes.

In table \ref{tab:U} we describe the parameters used for the
model sets, with the same inputs used for each metallicity. For each
input flux density and initial gas pressure we give the resulting
initial ionization state in terms of two parameters, $\Xi_{0}$ and
$\tilde{U}_{0}$ \citep{Groves06}. $\Xi_{0}$ is a measure of the
radiation pressure versus gas pressure ($\Xi_{0}=I_{0}/(cP_{gas})$), while
$\tilde{U}_{0}$ measures the relative number of ionizing photons $S_*$,
to a normalized initial density
($\tilde{U}_{0}=S_*/(c\tilde{n}_{\rm H})$, $\tilde{n}_{\rm H}=P_{\rm
gas}/(k10^4)$). 
 
The parameters explored here span the range in ionization
conditions that are observed in typical narrow line region clouds.
The models are all truncated at a column density of $\log 
(N$(\hi)$)=21$, which is reasonable for NLR clouds 
\citep[e.g.][]{Crenshaw03}.

\begin{table}
\caption{Input parameters for the models. The total incident flux
density $I_{0}$ is scaled by $38.098$\flux, while $P_{0 {\rm gas}}/k$ is
scaled by $10^6$K\pccm.}\label{tab:U}
\begin{center}
\begin{tabular}{|c|c|c|c|c|}
\hline
$model \#$ & $I_{0}$ & $P_{0 {\rm gas}}/k$ & $\Xi_{0}$ & $\log
(\tilde{U}_{0})$ \\
\hline
1 & 4.0 & 1.00 & 20.16    & -0.40 \\
2 & 3.0 & 3.20 & \phn4.73 & -1.03 \\
3 & 2.0 & 5.40 & \phn1.87 & -1.43 \\
4 & 1.0 & 7.60 & \phn0.66 & -1.88 \\
5 & 0.5 & 8.70 & \phn0.29 & -2.24 \\
6 & 0.25& 9.25 & \phn0.14 & -2.57 \\
7 & 0.1 & 9.58 & \phn0.05 & -2.98 \\
\hline
\end{tabular}
\end{center}
\end{table}

\section{Metallicity Effects on the Emission Lines}

When the total metallicity of a photoionized nebula decreases, there
are several effects that cause changes in the
final emission line spectrum. Many of these effects have been
described elsewhere \citep[eg][]{Osterbrock89,ADU03, Kewley02}. 

The dominant  effect is the decreased abundance of an emitting species
relative to hydrogen. This leads to 
weaker emission in lines such as nitrogen and  oxygen relative to
hydrogen recombination lines such as H$\alpha$ and H$\beta$.
The line ratios relative to the other abundant
primordial element, helium, also change in a similar way. This effect is 
not as strong, because helium also decreases with metallicity.

\begin{figure}
\includegraphics[width=\hsize]{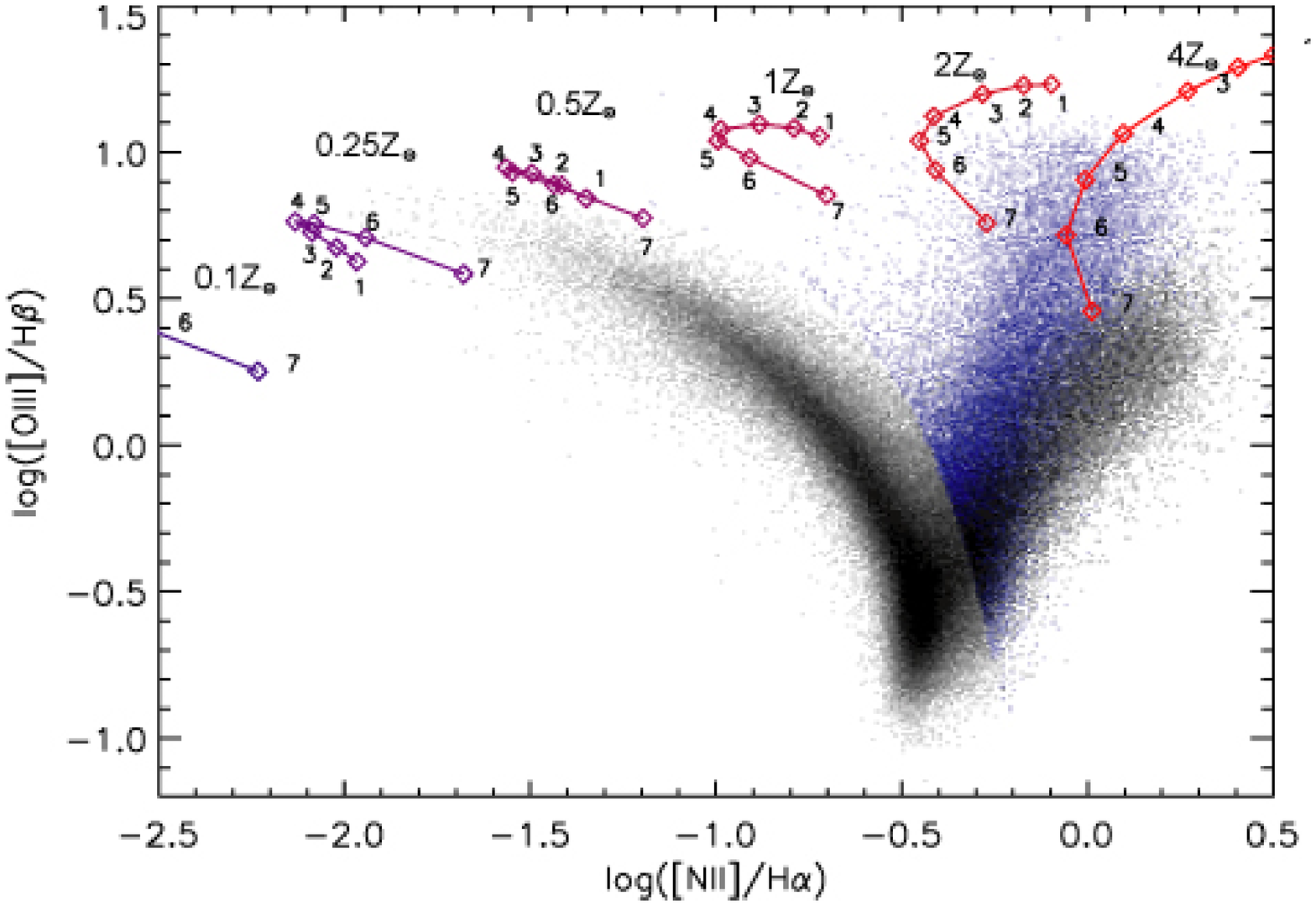}
\caption{The BPT diagram of [\nii]/\ha\ versus [\oiii]/\hb\
demonstrating the decrease of metal emission lines relative to
hydrogen. This decrease is stronger for the nitrogen lines due to the secondary
component in this element. Model metallicities are as labelled, with
the small numbers  
next to model indicating model run and associated ionization parameter
as given in table \ref{tab:U}. The background image shows the log scale
density distribution of the SDSS emission line sample, with Seyfert
galaxies lying
in the right, blue colored branch. Each branch has been
scaled separately to emphasize the position of the Seyfert galaxies.
}\label{fig:nii}
\end{figure}

At metallicities above $0.1Z_{\odot}$, nitrogen is dominated by
its secondary component. A decrease in  metallicity
down to this value results in a decrease of 
the strength of the nitrogen emission line relative to  
that of other metals. 
Below $Z \sim 0.1Z_{\odot}$, the primary component
is dominant and the nitrogen tracks the behaviour of the
other metal lines more closely.
These trends are clearly seen in figure \ref{fig:nii}, which
presents the models on the classic \citet{BPT81} (BPT) diagram of
[\nii]$\lambda 
6584$\AA/\ha\ versus [\oiii]$\lambda 5007$\AA/\hb. This diagram clearly reveals
the strong decrease of the nitrogen emission lines relative to
hydrogen as the metallicity decreases. 

As metallicity decreases, the temperature of the nebula also increases.
This is a result of the decrease in the efficiently cooling metal
emission lines and the availability of more high energy photons
to ionize hydrogen \citep[see][ for detailed explanations]{Sutherland93}. 
Higher nebular temperatures affect the strength of temperature
sensitive emission lines, such as [\oiii]$\lambda 5007$\AA, which
increases with temperature. Line 
ratios like [\oiii]$\lambda 5007$\AA/\hb\ will thus decrease at a
relatively slower rate with metallicity. This is also clear from
the model presented in Figure 2. 
While temperature sensitive
line ratios
change with the metallicity of the gas, they are also 
strongly affected by the ionization
state of the gas. This means that shocks and the varying contribution of
star formation and AGN to the ionization  become important.

Another effect of lowering the metallicity is that the opacity of the
gas is reduced, meaning a greater volume is needed to absorb all
ionizing photons. 
In AGN this effect causes noticeable differences because of the
stronger contribution 
to the ionizing flux from X-rays, for which
the hydrogen opacity is small. These X-rays give rise to the partially
ionized region in NLRs, where important optical emission lines such as
[\oi]6300\AA\ and [\sii]6713+31\AA\ are emitted.
Models with lower metallicity have larger 
partially ionized regions relative to the fully
ionized region \citep[see][]{Groves04a}. 
In our column density limited models, the partially ionized region 
becomes  truncated at low metallicities. 
As a result, we are not confident in our
ability to use these models to 
calibrate some of the metallicity sensitive ratios commonly used
in determining the abundances in star-forming galaxies, such as
[\nii]/[\sii] and [\sii]/\ha\ for use in the determination of AGN abundances.

\section{Diagnostic Diagrams for Low Metallicity AGN}\label{sec:diag}

Generally, individual AGN are best reproduced with a combination of
models, varying not only the metallicity, but also
the ionizing spectrum, the density, the incident flux or ionization
parameter and even the geometry 
\citep[see eg][]{Oliva99}.

However when looking at large numbers of AGN the best method is to
look for trends or possible
relationships between the line ratios and metallicity
\citep[eg][]{Storchi98,Nagao02,Groves04b}. Previous AGN modelling has
found that  
AGN as a group are best fit with models of around $2-4Z_{\odot}$, or
$12+\log($O/H$)\sim 9.0-9.3$. Thus any AGN NLR found to have
metallicities below solar could possibly be considered ``low metallicity''.

As discussed in the previous section, line ratios relative to hydrogen
are good diagnostics for low metallicity AGN. 
This is especially true for nitrogen with
its secondary component as shown in figure \ref{fig:nii}. The
diagnostic of [\nii]$\lambda 6584$\AA/\ha\ has been used as one of the
main metallicity indicators in AGN due to its
sensitivity and the strength of the lines \citep{Storchi89,Storchi98}.  
In general, low
metallicity AGN will lie to the left of the Seyfert branch on the
\nii\ BPT diagram. The decrease in the [\oiii]/\hb\ ratio is not as
strong because of the temperature sensitivity of this ratio.

\begin{figure}
\includegraphics[width=\hsize]{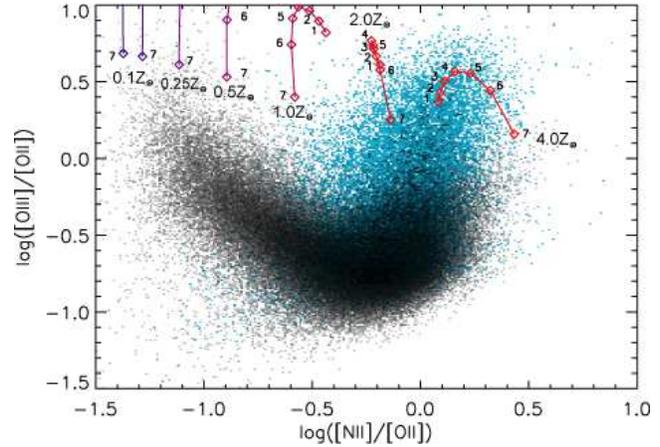}
\caption{Metallicity diagnostic diagram of [\nii]$\lambda
6584$\AA/[\oii]$\lambda 3727$\AA\ versus [\oiii]$\lambda 5007$\AA/
[\oii]$\lambda 3727$\AA\ with models increasing in metallicity from
left to right (as labelled). The background image shows the
distribution of the SDSS emission line galaxies, with Seyfert galaxies,
colored blue, lying in the vertical
branch and starbursts in the horizontal branch. Each branch has been
scaled separately to emphasize the position of the Seyfert galaxies.
The contribution by star formation to the AGN 
decreases the [\oiii]/[\oii] ratio.}\label{fig:oiioiii}
\end{figure}

Another possible diagram for metallicity diagnosis is the
[\nii]$\lambda 6584$\AA /[\oii]$\lambda 3727$\AA\ versus
[\oiii]$\lambda 5007$\AA/ [\oii]$\lambda 3727$\AA\ diagram, shown in
figure \ref{fig:oiioiii}. This again relies upon nitrogen to separate
the different metallicities, but also uses the temperature sensitive ratio 
[\oiii]/[\oii] as a secondary diagnostic. 
The [\nii]/[\oii] ratio will
separate AGN with lower \nii\ relative to \oii. As \nii\ and
\oii\ are both low ionization species , the ratio will only be weakly
affected by 
the ionization or density conditions.
In addition, star
bursts and AGN  have similar [\nii]/[\oii] ratios for similar metallicities, so
the abundance determination of an active galaxy is only slightly
perturbed by any 
contribution from star formation.
The main disadvantage of this ratio is that [\nii] and [\oii] are widely 
separated in wavelength and the ratio is hence much more sensitive to
the effects of 
reddening.
The [\oiii]/[\oii] ratio is similar to [\oiii]/\hb\ as a temperature
sensitive diagnostic in that is good in separating the Seyferts from the
star-forming galaxies. It also has the additional benefit of not being
strongly sensitive to metallicity. 

Both these diagnostic diagrams rely heavily on the
[\nii] line to estimate the metallicity. 
This means that the diagnostic is sensitive to abundance variations
relative to solar within the NLR due to differing star formation
histories. 

\begin{figure}
\includegraphics[width=\hsize]{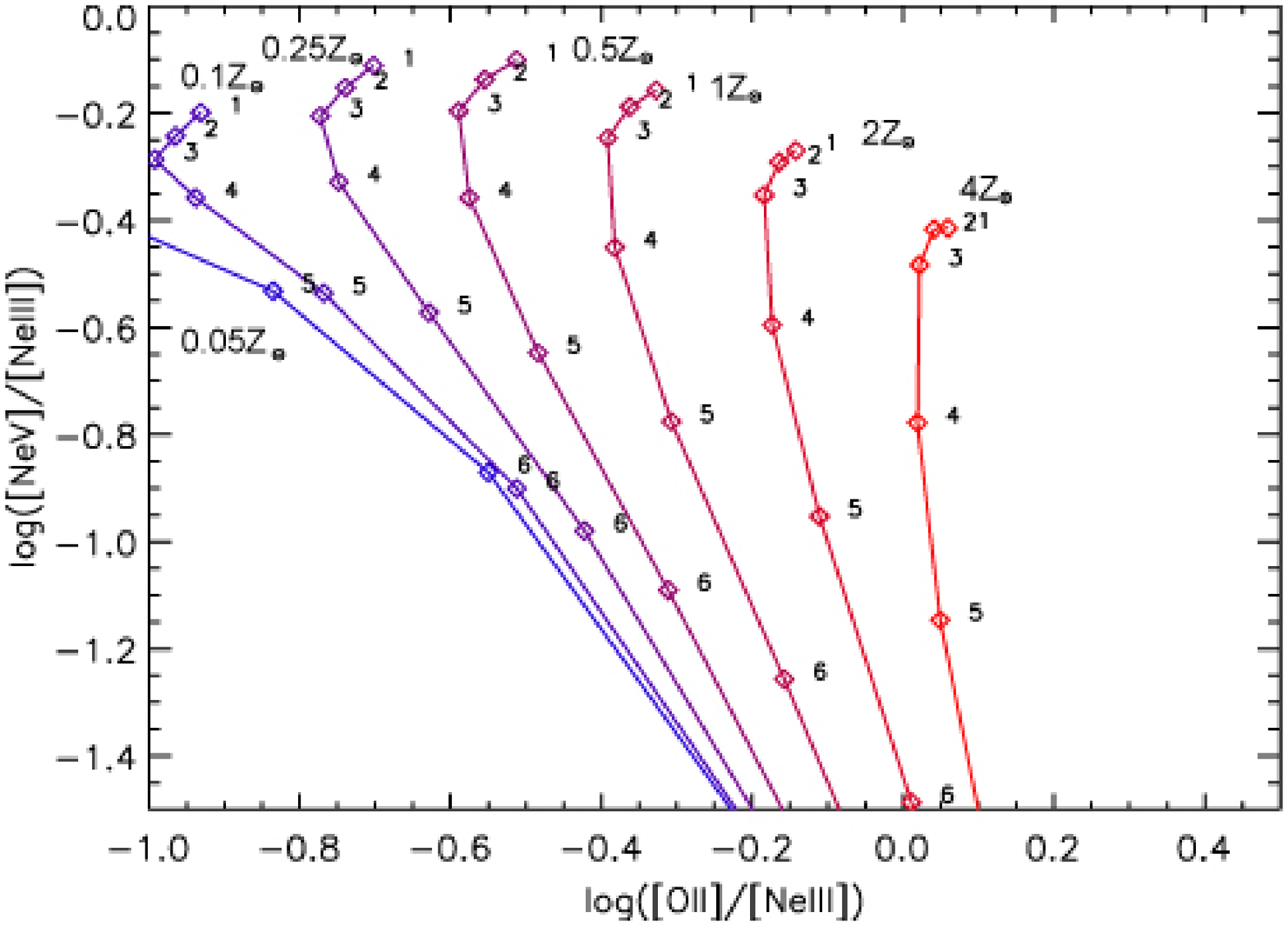}
\caption{Near--UV metallicity diagnostic diagram of
[\oii]$\lambda3727,9$\AA/[\neiii]$\lambda3869$\AA\ versus 
[\nev]$\lambda3426$\AA/[\neiii]$\lambda3869$\AA. Pure star forming
galaxies would lie below these curves at low [\nev]/[\neiii] values.
}\label{fig:UV}
\end{figure}

At higher redshift, the [\nii] line is no longer observable,
but UV diagnostics become useful 
\citep{Nagao06,Groves04b}. 
One possible diagram is shown in figure \ref{fig:UV}, which plots the
Near--UV line ratios of
[\oii]$\lambda3727,9$\AA\ /[\neiii]$\lambda3869$\AA\ versus
[\nev]$\lambda3426$\AA/[\neiii]$\lambda3869$\AA. This diagram relies upon 
the temperature sensitivity of the [\neiii] line to distinguish
different metallicities in the [\oii]/[\neiii] ratio. The
[\nev]/[\neiii] ratio, while weakly dependent upon metallicity, is
very sensitive to the ionizing spectrum and ionization state of the
gas. Thus this ratio provides a way of distinguishing
star-forming galaxies, which have very little [\nev] emission, from AGN. 
As in figure \ref{fig:oiioiii}, the strength of figure \ref{fig:UV}
lies in its use of strong lines and its ability to distinguish 
star forming galaxies from AGN. 

\section{AGN Selection within SDSS}

Within this work we use the SDSS Data Release 4 (DR4) spectroscopic galaxy
sample \citep{DR4}, which includes $u$-, $g$-, $r$-, $i$- and $z$-band
photometry and spectroscopy of over 500,000 objects. 
The spectra are taken using
3-arcsec diameter fibres, positioned as close as possible to the
centres of the target galaxies. The flux- and wavelength-calibrated
spectra cover the range from 3800 to 9200\AA, with a resolution of
$R \sim$1800. 

At the median redshift of the sample ($z\sim 0.1$) the spectroscopic
fibre typically contains 20 to 40 percent of the total galaxy light,
thus contain a component due to the host galaxy as well
as the AGN. 
As described in \citet{Tremonti04}, we subtract 
the contribution of the stellar continuum from each spectrum, using the
best fitting 
combination of template spectra from the population synthesis code of
\citet{Bruzual03}. The best-fitting model also places constraints on the
star formation history and metallicity of the
galaxy\citep{Gallazzi05}, 
and can be used to estimate
stellar masses and star-formation histories \citep{Kauffmann03a}.

We have used the same sample criteria as in                      
\citet{Kewley06} to extract our narrow-line  AGN subsample.
The sample is limited to redshifts above 0.02.               
We also apply a signal-to-noise cut ($S/N > 3$) on the
six dominant emission lines; 
\hb\ $\lambda 4861$\AA, [\oiii] $\lambda
5007$\AA, [\oi] $\lambda 6300$\AA, \ha\ $\lambda 6563$\AA, 
[\nii] $\lambda 6584$\AA, and the doublet [\sii] $\lambda 6716,31$\AA. 
This gives a sample of $\sim 170,000$ emission line galaxies.
The \citet{Kauffmann03c} \nii\ empirical relation is then used to
define our AGN sample, 
\begin{equation}\label{eqn:nii}
\log([\mathrm{OIII}]/\mathrm{H}\beta) >
0.61/(\mathrm{NII}]/\mathrm{H}\alpha-0.05)) + 1.3.
\end{equation}
This results in a sub-sample of $\sim 50000$ galaxies classified as AGN.
The \nii\ dividing line of \citet{Kewley02},
\begin{equation}\label{eqn:niiK}
\log([\mathrm{OIII}]/\mathrm{H}\beta) >
0.61/(\mathrm{NII}]/\mathrm{H}\alpha-0.47)) + 1.19,
\end{equation}
is used to select a subsample of 20,000 ``pure'' AGN where the
ionization is dominated  
by the active nucleus rather than \hii\ regions.

The AGN sample includes two classes of emission line
objects; Seyfert 2s and Low Ionization Nuclear Emission-line Regions
(LINERs). 
In this work we consider only Seyfert galaxies, as our 
diagnostic models for the NLR only  apply to these objects. 
To distinguish LINERs  from Seyferts, we use the recent
empirical dividing lines from \citet{Kewley06}
\begin{equation}\label{eqn:oiL}
\log([\mathrm{OIII}]/\mathrm{H}\beta) >
1.36\log([\mathrm{OI}]/\mathrm{H}\alpha) + 1.4,
\end{equation}

Our final sample consists of 8800 pure Seyfert galaxies
ranging in redshift 
from $z\sim 0.02$ to $\sim 0.36$, with a median of $z\sim 0.1$.

\subsection{Accounting for Reddening in the SDSS Sample}\label{sec:red}

Line ratios such as  [\nii]/[\oii] have a large separation
in wavelength and thus need to be corrected for reddening.             
We use the  \ha/\hb\ ratio and we assume dust-free values of 
2.86 for star-forming galaxies and 3.1 for Seyferts. The majority of
AGN in our sample will  have ongoing starformation and their dust-free
values of this ratio should lie between these values.
We then assume the power-law
slope of $\lambda^{-0.7}$ from \citet{Charlot00} 
to correct for the
attenuation in each galaxy.
 
\section{Low Metallicity AGN in SDSS}

Measurements of the average gas metallicities in starforming galaxies
\citep[fig.~6,][]{Tremonti04} show that there is a strong trend of
increasing metallicity with increasing stellar mass. This trend is
also seen in the average stellar metallicities
in normal galaxies \citep[fig.~8,][]{Gallazzi05}.

\begin{figure}
\includegraphics[width=\hsize]{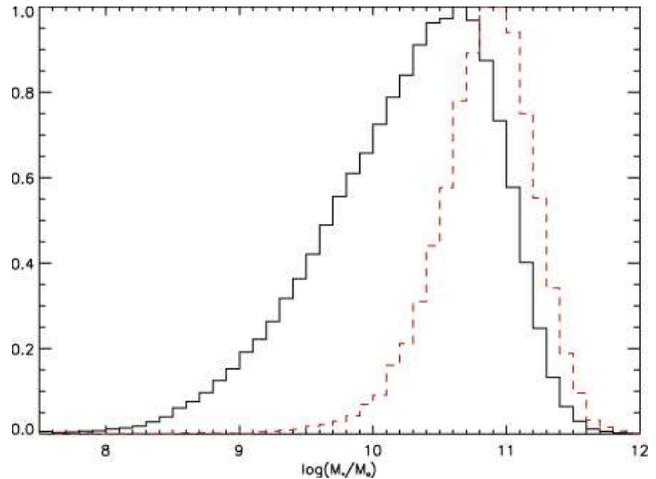}
\caption{The Stellar Mass distribution of the full emission line
galaxy sample (\emph{solid line}) and the AGN (including LINER) 
dominated galaxies
(\emph{dashed line}) in Solar masses. 
The peaks have been scaled by 11013 and 2286
for the full and AGN sample respectively.}\label{fig:Masshist}
\end{figure}

Figure \ref{fig:Masshist} shows the mass distribution of the full
emission-line galaxy sample and the AGN sample. AGN are clearly biased
towards higher masses
\citep{Kauffmann03c} and they range   
from $\sim 10^{10}$ to $\sim
10^{11.5}$M$_{\odot}$. If AGN have similar metallicities
to star-forming galaxies, they should have abundances in the range 
$12+\log({\rm O/H})\sim 8.95$ to $\sim 9.13$ \citep[][
eqn.~3]{Tremonti04}, or metallicities of 
2-3$Z_{\odot}$. As shown in figure \ref{fig:nii} this is also the 
metallicity range for which  our AGN  photoionization models
best reproduce the emission line ratios of the typical Seyfert
galaxies within our sample.

\begin{figure}
\includegraphics[width=\hsize]{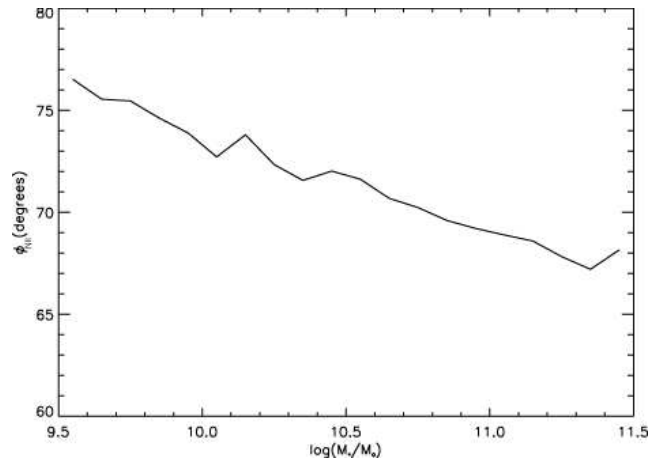}
\caption{Variation of the median angle of the Seyfert branch,
$\phi_{NII}$, 
with the host stellar mass, $M_{*}$
in solar masses. $\phi_{\mathrm{NII}}$ is the angle relative to the $x$-axis
on the BPT [\nii] diagram using the empirical point $[-0.5,-0.6]$ as
the vertex (see figure \ref{fig:niilowM}). Only AGN-dominated Seyfert
classified galaxies are considered. 
}\label{fig:mass-phi}
\end{figure}

\begin{figure}
\includegraphics[width=\hsize]{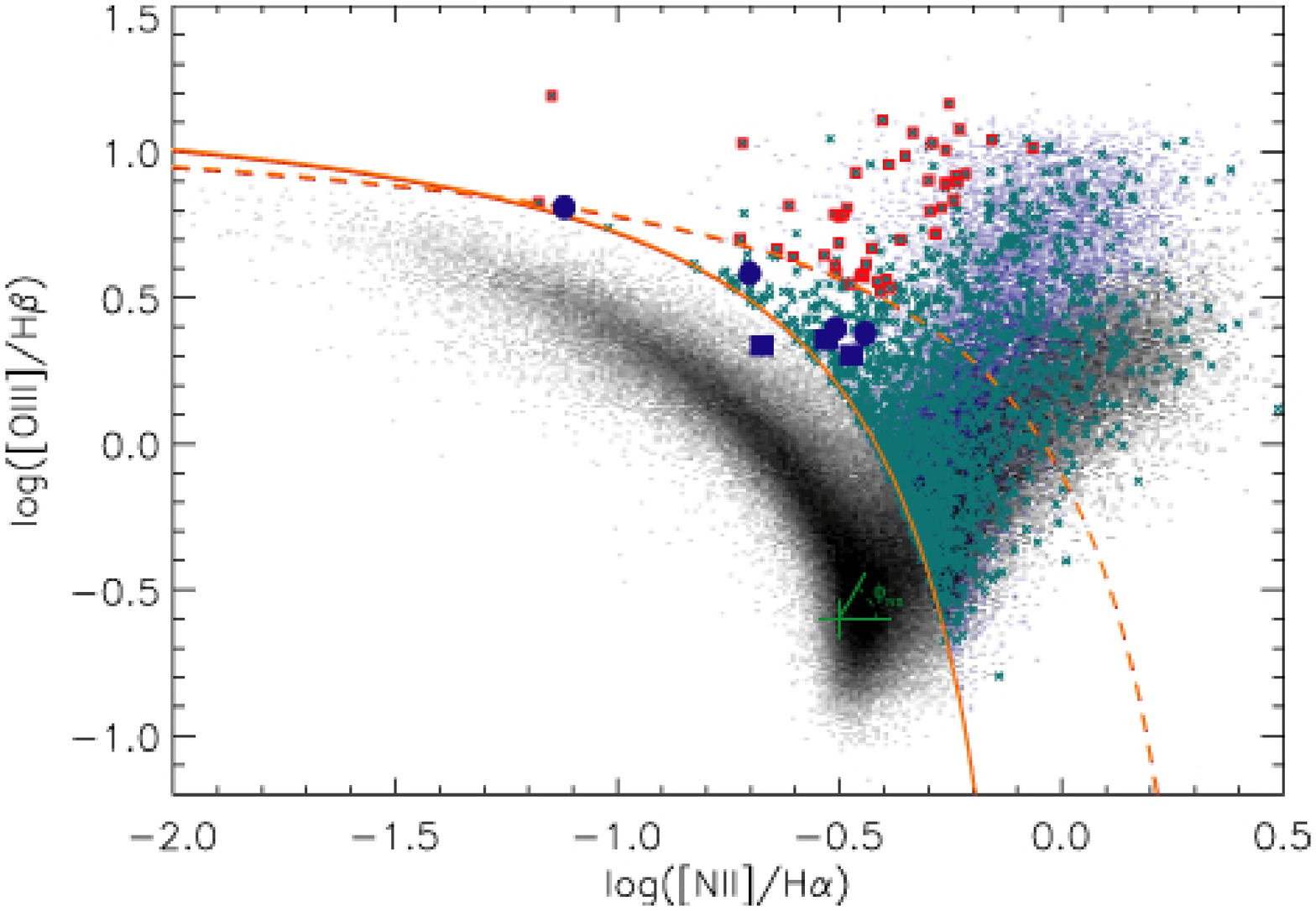}
\caption{The BPT [\nii]/\ha\ versus [\oiii]/\hb\ diagram showing the
distribution of the low mass AGN (\emph{crosses}). The 
background image shows the log scale density distribution of the full
SDSS emission line sample as in figure \ref{fig:nii}. The solid and
dashed curves show the Kauffmann and Kewley classification lines
respectively. The boxed crosses highlight the lowest metallicity low
mass Seyfert galaxies (see figure \ref{fig:lowMmetal}).
The solid squares and circles represent the \citet{Shapley05} and
\citet{Erb06} high redshift 
star-forming galaxies respectively, and show their position relative to the low
metallicity AGN. (see \S \ref{sec:discuss}).
}\label{fig:niilowM}
\end{figure}

Figure \ref{fig:nii} shows that as the metallicity decreases, AGN are
expected to 
migrate to the left on the [\oiii]/\hb\ versus [\nii]/\ha\
BPT diagram. 
Following \citet{Kewley06}, we choose an empirical base
point on the [\nii] BPT diagram ([-0.5,-0.6], as marked on figure
\ref{fig:niilowM}) and use this as the vertex to determine the angle
each AGN in our sample makes with the $x$-axis. We label this angle
$\phi_{\mathrm{NII}}$. 
Figure \ref{fig:mass-phi} shows
the variation of the median value of this angle 
as a function of the stellar mass of
the host. 
This figure clearly reveals a correlation between the host stellar
mass and metallicity. 
The change in $\phi_{NII}$ from $\log M_* \sim 9.5$ to $\log M_* \sim 11.5$
corresponds to a change of about 0.2 -- 0.3 dex in log(O/H) in the AGN
models.  

We then use a cut in stellar mass to separate out a sample 
of low metallicity AGN for closer inspection.
We select AGN hosts with  log(M$_{*}/\mathrm{M}_{\odot})<10.0$. 
This limit selects the
tail end of the AGN mass distribution, while maintaining sufficient
numbers for analysis. This yields 1800 low-mass AGN, of which  650 lie
above the Kewley line.
Figure \ref{fig:niilowM}  shows the
distribution of the low-mass  AGN (both Seyferts and
LINERs) on 
the standard BPT [\nii] diagram. As can be seen, the low mass objects, while
scattered, tend to lie towards the left of the AGN branch.
Many of them are strongly clustered close to the Kauffmann dividing line.
To reduce the effects of confusion and  contamination by star-formation,
we concentrate on the Seyfert galaxies that lie above the Kewley line.
This reduces the sample to $\sim 350$ galaxies, 

\begin{figure}
\includegraphics[width=\hsize]{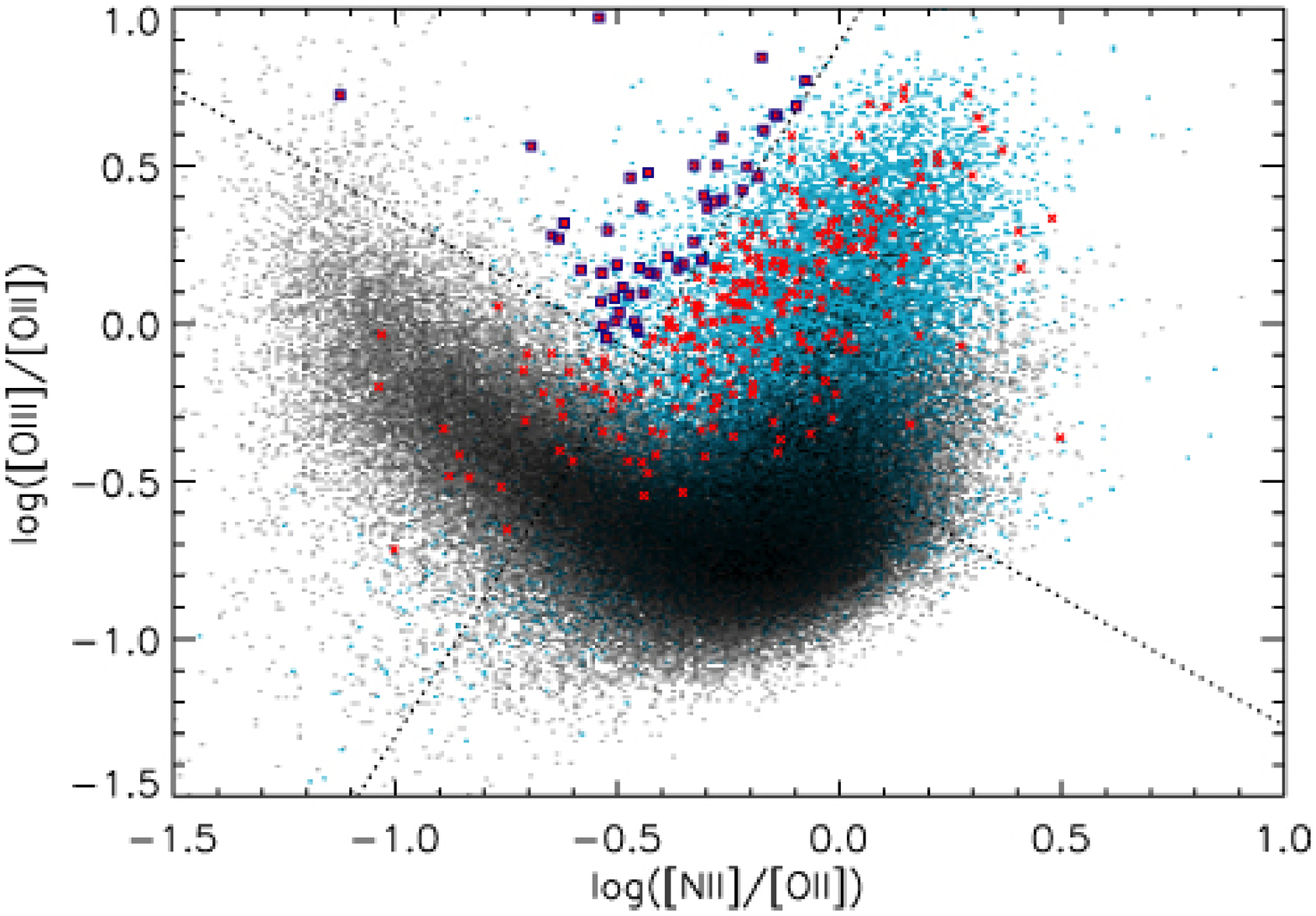}
\caption{The metallicity sensitive [\nii]$\lambda
6584$\AA/[\oii]$\lambda 3727$\AA\ versus 
[\oiii]$\lambda 5007$\AA/[\oii]$\lambda 3727$\AA\ diagram showing the
distribution of the low mass Kewley-selected AGN galaxies
(\emph{crosses}). The
background image shows the logscale density distribution of the full
SDSS emission line sample. The Seyfert galaxies, marked as blue on the
image, lie on the vertical branch at high
[\nii]/[\oii], while the star-forming galaxies lie on the mainly
horizontal curve. LINERs exist at the dark clump on the star-forming
branch at high [\nii]/[\oii].
 The dotted lines indicate our arbitrary metal \& AGN
cuts, with our final low mass, low metallicity AGN highlighted by
squares.}\label{fig:lowMmetal}
\end{figure}

To estimate the metallicity of the low mass AGN we plot these objects in the
[\nii]$\lambda 6584$\AA/[\oii]$\lambda 3727$\AA\ versus
[\oiii]$\lambda 5007$\AA/[\oii]$\lambda 3727$\AA\ diagram described
in the previous section. This is shown in figure
\ref{fig:lowMmetal}. The use of the Kewley cut means    
that the majority of starburst dominated AGN have been removed from
the sample as can be seen from the scarcity of objects in the
heavily shaded greyscale region. However there are still some objects which
appear to lie in the starburst branch. These galaxies may either have
incorrectly measured line fluxes or incorrect reddening corrections (as
is certainly the case for  the lowest [\oiii]/[\oii] object).
Once again the low mass Seyferts tend to lie to the left, low
metallicity side of 
the main Seyfert branch.

Figure 8 demonstrates is that even at the low mass end of the
AGN distribution, the AGN gas metallicity is still around solar
(accounting for dust depletion). We note that this is similar to the
gas-phase  metallicities of star-forming galaxies of the same mass        
\citep[equation 3 in][]{Tremonti04}.

To select the most extreme 
 low metallicity AGN in our sample, we apply two
cuts to exclude those objects that lie within
the main AGN or starforming branches. These cuts are shown by the dotted
lines in figure \ref{fig:lowMmetal} and are:
\begin{equation}
\begin{array}{rcl}
\log([\mathrm{OIII}]/[\mathrm{OII}]) &\ge & 
2.1\log([\mathrm{NII}]/[\mathrm{OII}])+0.85\\
\log([\mathrm{OIII}]/[\mathrm{OII}]) &\ge & 
-1.2\log([\mathrm{NII}]/[\mathrm{OII}])-0.7.
\end{array}
\end{equation}

These cuts select 50 of the 559 low mass AGN as our lowest
metallicity candidates, and they are plotted as boxed crosses on figures
\ref{fig:niilowM} and  \ref{fig:lowMmetal}. Note that one object has
multiple spectra, which  
leaves 47 unique candidates. Six of these objects lie at redshifts
greater than 0.1, with at least one 
being a strongly starbursting galaxy.
 These low ``metallicity'' AGN lie clearly to
the left of the AGN branch in the BPT [\nii] diagram (see figure
\ref{fig:niilowM}), and similarly lie on the left part of the AGN
branch in both
the BPT [\sii] and [\oi] diagrams. The few low mass AGN that lie to
the left of the AGN branch on figure \ref{fig:niilowM} and are not
selected by our cuts are actually at too low a redshift ($z
\lapprox0.03$) for accurate measurement of the [\oii] line flux.

\onecolumn
\begin{figure}
\includegraphics[width=0.48\hsize]{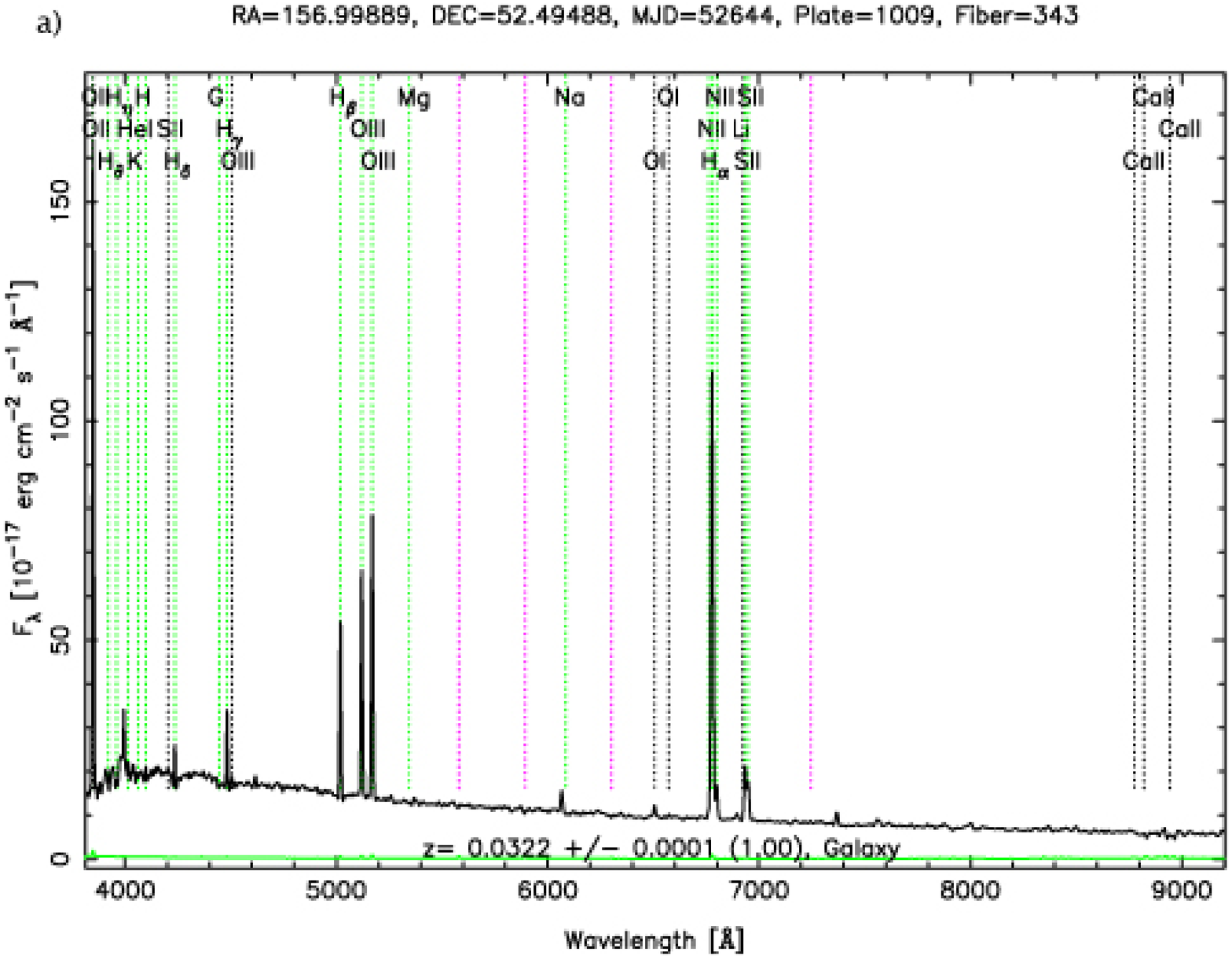}
\includegraphics[width=0.48\hsize]{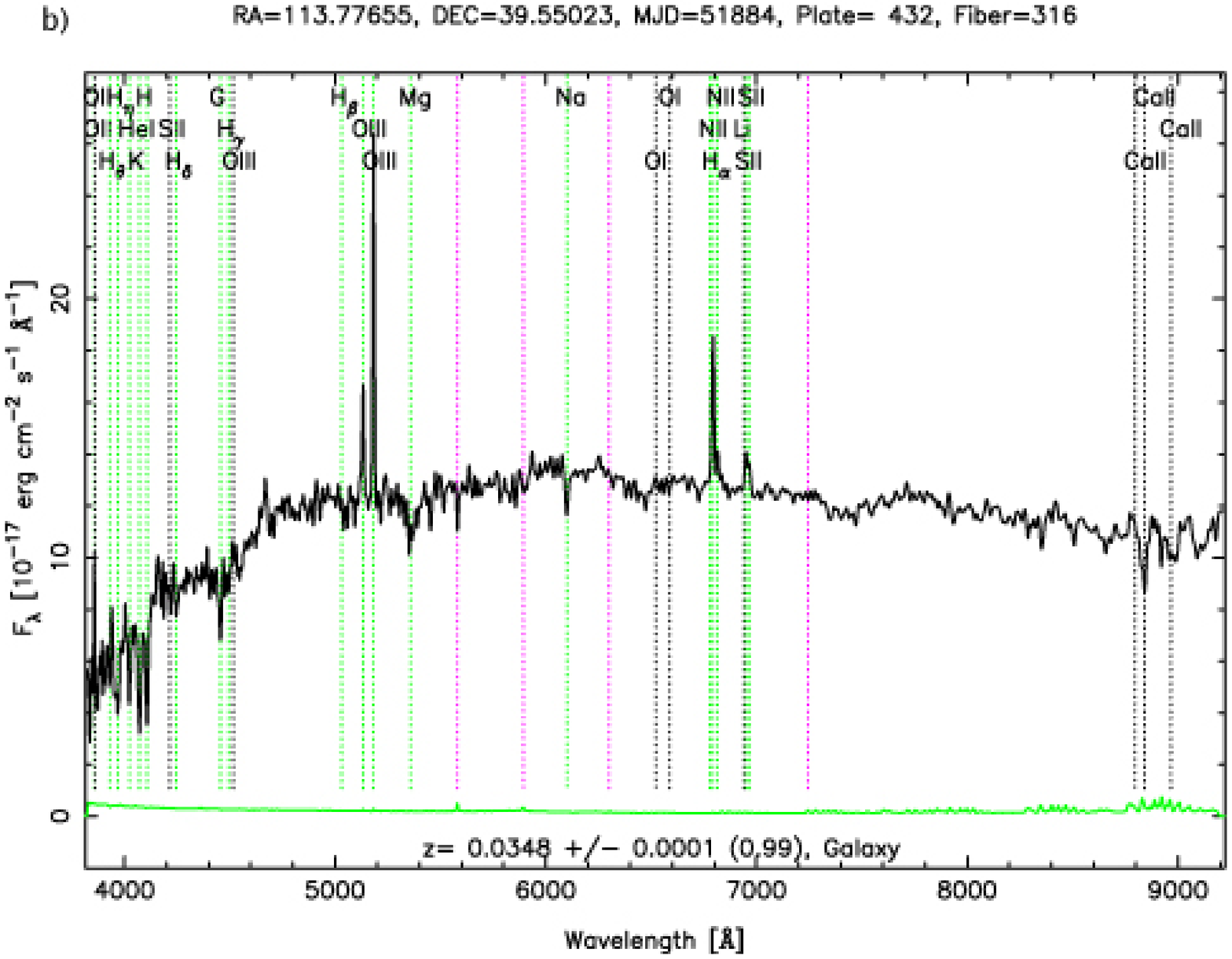}\\
\vspace{5mm}
\includegraphics[width=0.48\hsize]{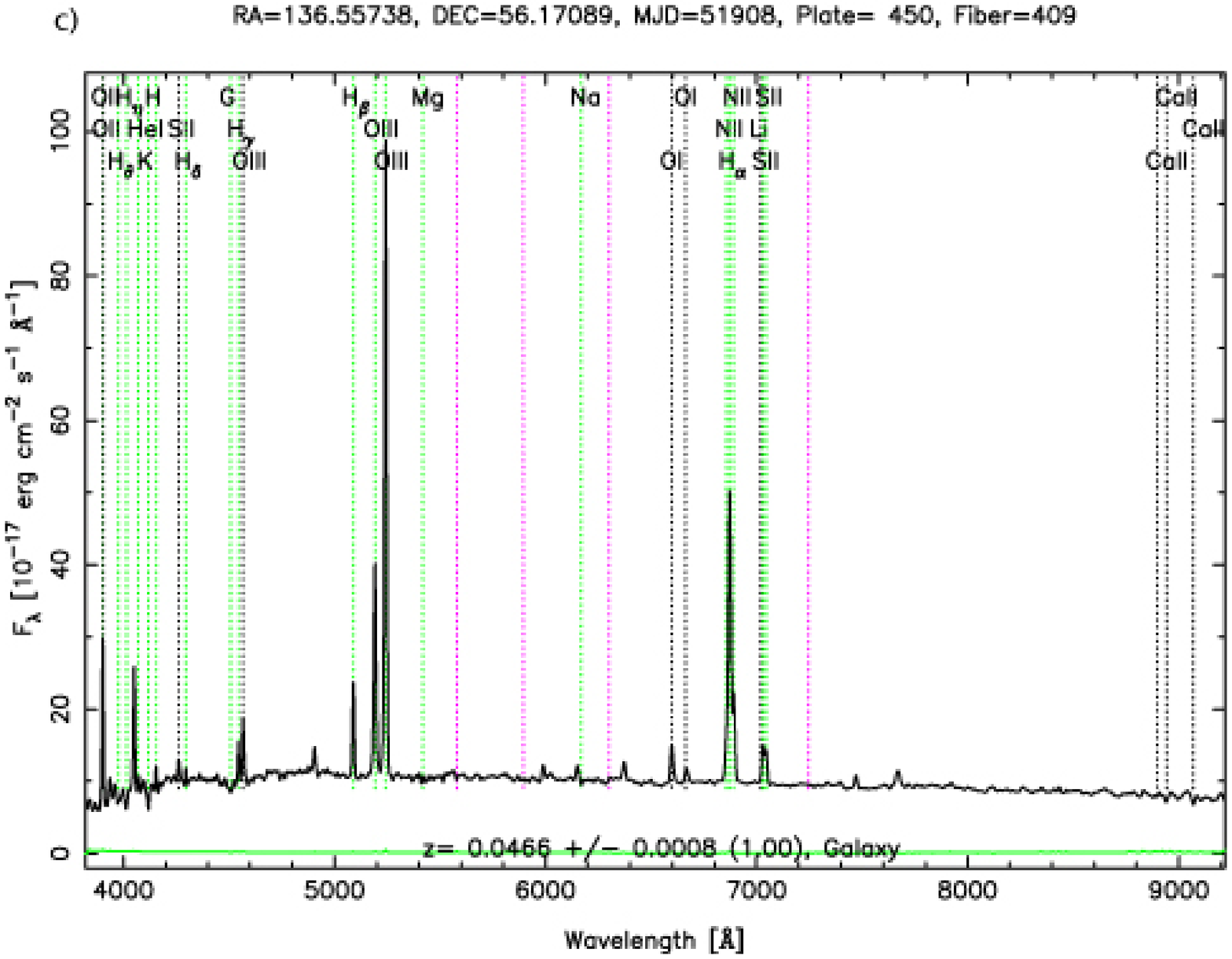}
\includegraphics[width=0.48\hsize]{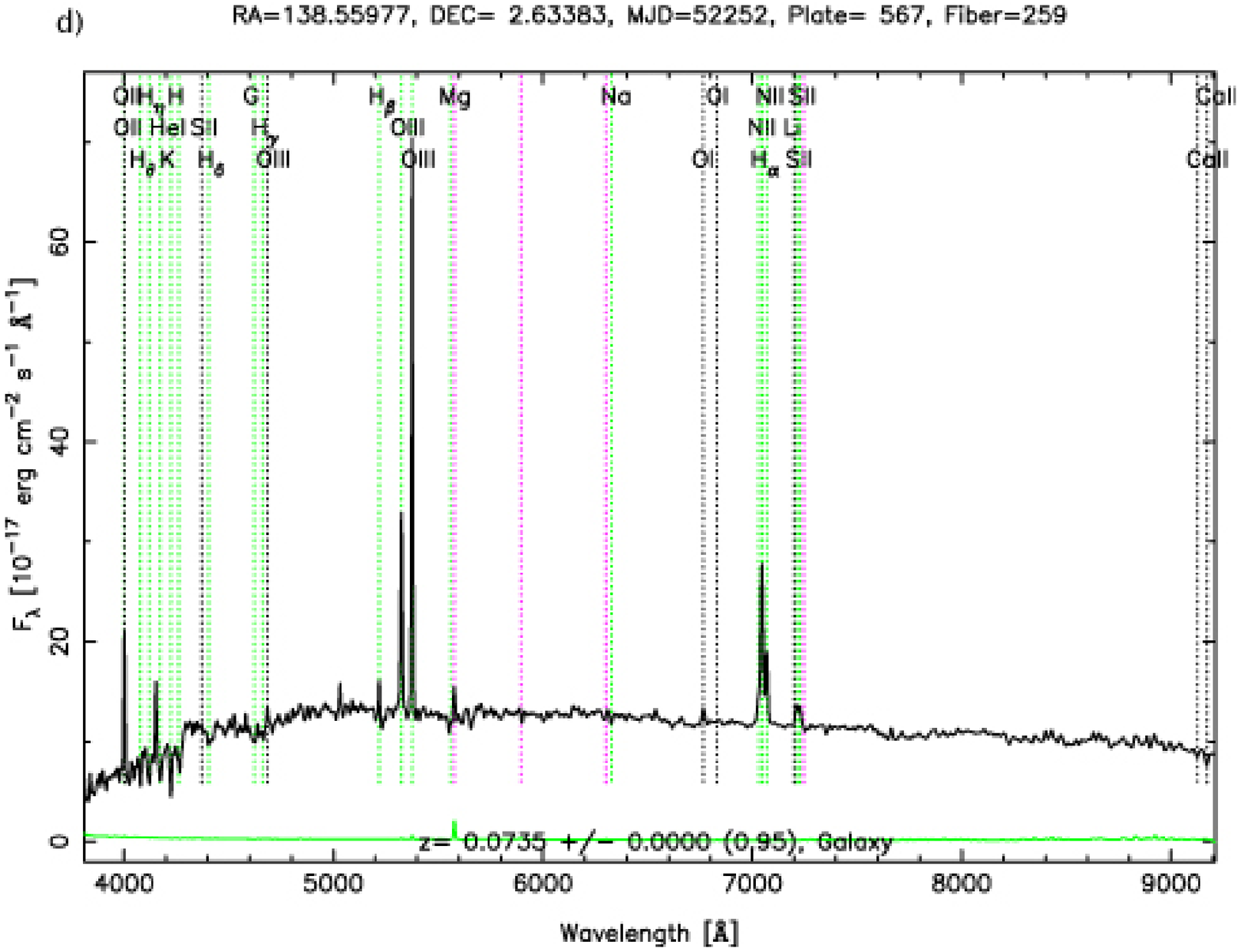}\\
\vspace{5mm}
\includegraphics[width=0.48\hsize]{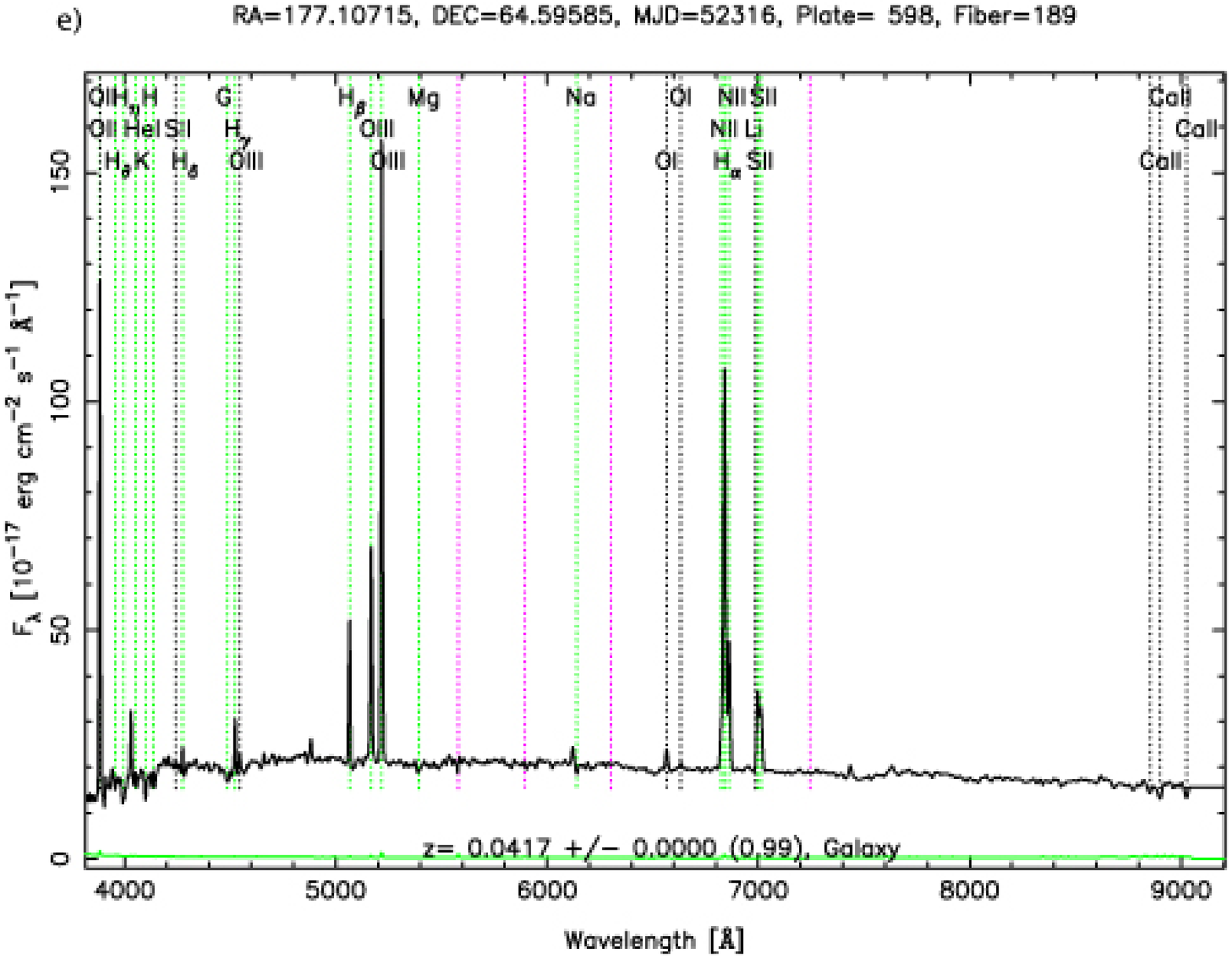}
\includegraphics[width=0.48\hsize]{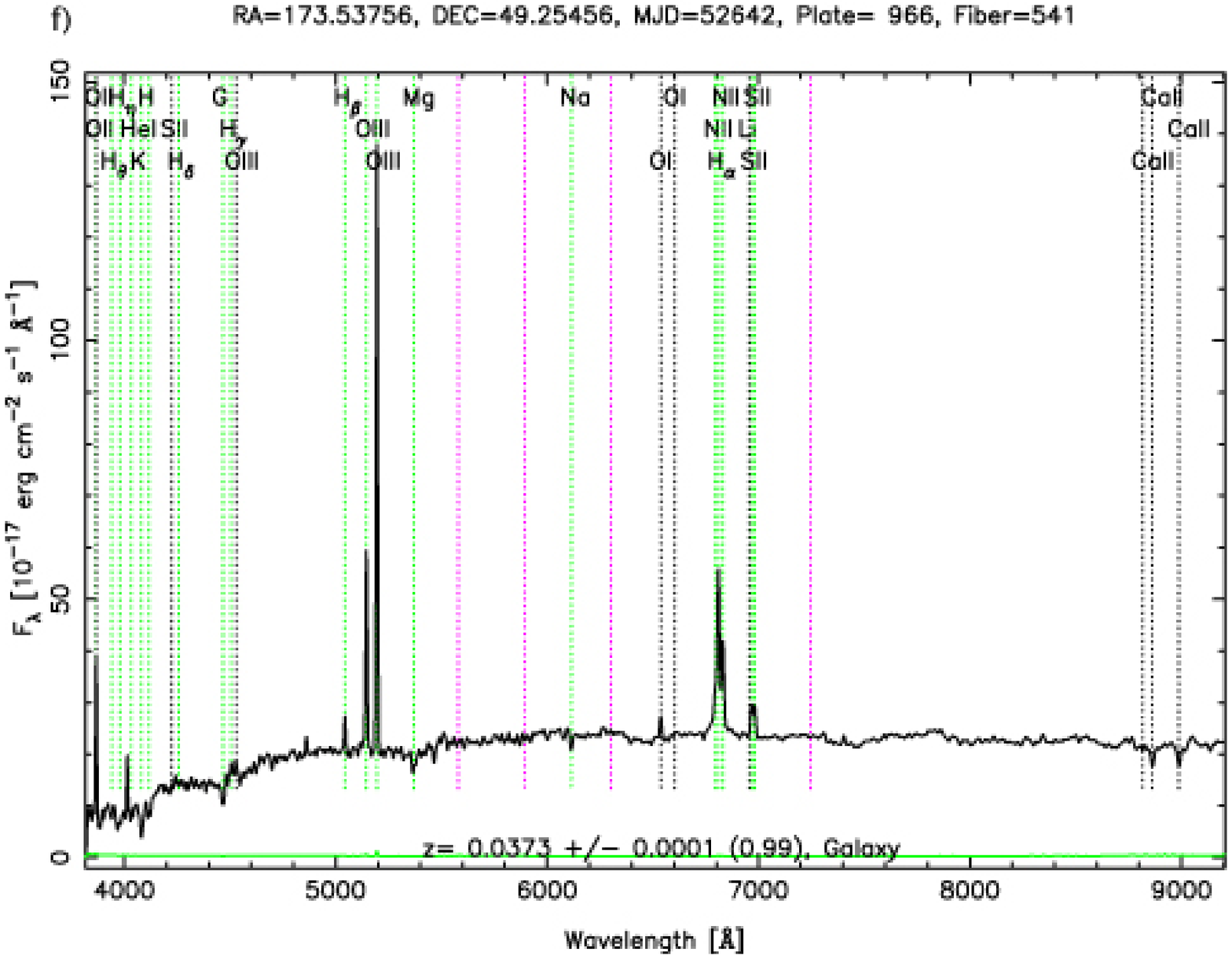}
\caption{SDSS fibre spectra of six of the low mass, low
metallicity selected AGN. a) shows the spectrum of a 
starburst galaxy that falls in our selection criteria
 while b) -- f) shows five representative low
metallicity AGN. The difference between the starburst and AGN can be
seen, as well as other AGN signatures, such as the strong [\heii]$\lambda
4686$\AA\ line (unmarked line to the left of \hb), or old stellar
continuum.}\label{fig:lowMspec}
\end{figure}
\twocolumn
As a final check we examined the spectra of these objects and a       
representative selection of these is
shown in figure \ref{fig:lowMspec}. One of these objects (figure
\ref{fig:lowMspec}a), the galaxy 
with the lowest [\nii]/[\oii] ratio, is a clear starforming 
galaxy with an apparent incorrectly estimated [\oiii]5007
strength. However this is the only
clear pure starburst galaxy among the 41 
objects. 
Some of the spectra of the other 40 AGN candidates exhibit clear
signatures of young stars as well as \heii\ $\lambda 4686$\AA\
emission.
Others are dominated by older stars, and there are also a number of
AGN
with strong 
Balmer absorption lines. All objects have very  
weak [\nii]6584 and [\oi]6300 (expected because of their low metallicities),
but the strong [\oiii]5007 emission line 
and relatively strong [\sii] 6716,31 doublet
classify these objects as AGN.
One other point to note is the presence of the [\oiii]$\lambda
4363$\AA\ line in many of the spectra, indicative of the high gas
temperatures in these objects, and also suggestive of low
metallicities. 

\begin{figure}
\centering
\includegraphics[width=\hsize]{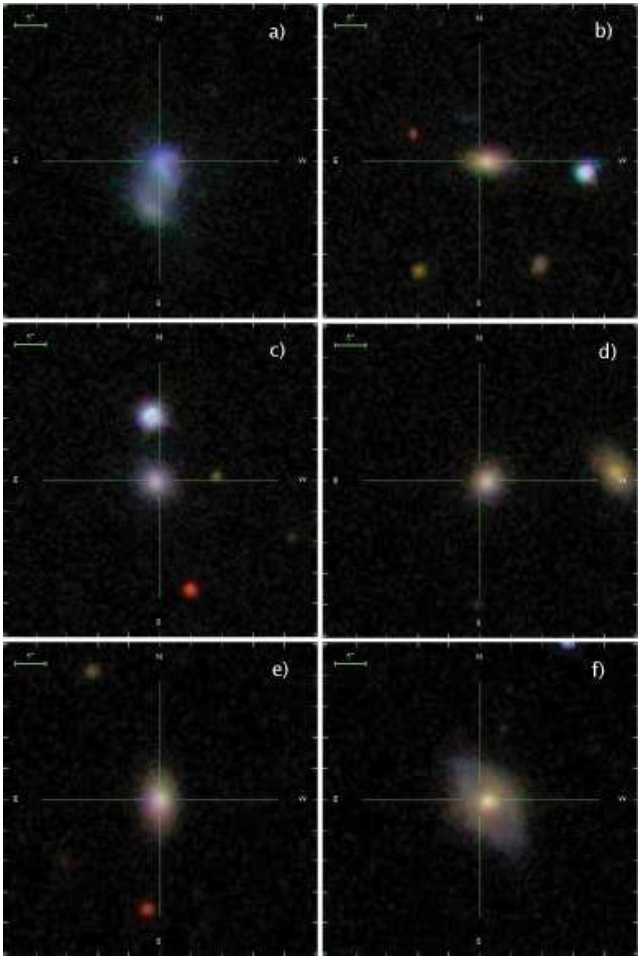}
\caption{SDSS images of the six sample galaxies from figure
\ref{fig:lowMspec} with a) an example starburst galaxy and b) -- f) low mass
AGN. Each image shows the same area, with the bar in
each frame indicating 5\mbox{\arcsec}. The image centre, as marked by the
lines, is 
the position of the 3\mbox{\arcsec} SDSS fibre centre. 
Note that for most AGN the
3\mbox{\arcsec} fibre encompasses a significant fraction of the host galaxy. 
}\label{fig:lowMimg}
\end{figure}

\begin{figure}
\centering
\includegraphics[width=\hsize]{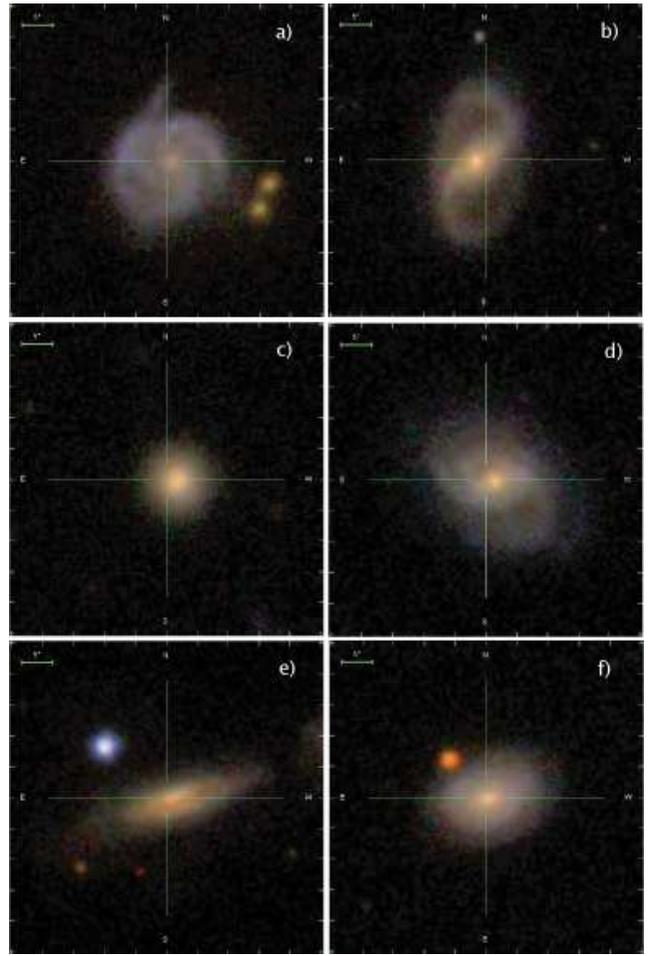}
\caption{SDSS images of six random NLR dominated Seyfert
galaxies. These galaxies have redshifts $z=0.05\pm0.01$, and all fall
within the main Seyfert branch on line diagnostic diagrams. Markings
the same as in figure \ref{fig:lowMimg}.
}\label{fig:Seyimg}
\end{figure}

As a complement to the SDSS fibre spectra, we also show images of the
same  sample in figure \ref{fig:lowMimg}. 
The starforming galaxy has a distinctly different morphology compared
to  the AGN, appearing 
diffuse and blue in color.
The images of the low mass AGN indicate that their host galaxies are
faint and   
relatively compact. They tend to be red, with a disky or lenticular structure 
surrounding a relatively bright nucleus.
For comparison, in figure \ref{fig:Seyimg}
 we show Seyfert galaxies that fall in the main branch on line diagnostic
diagrams and have a median
redshift similar to the low mass sample of $z\sim 0.05$. The most
obvious difference is their  size. The spiral structure 
in the host galaxy is also much more evident. However in some
respects the low mass and high mass  AGN are similar; both have a       
bright red nucleus/bulge, and diffuse, blue outer disk.

\subsection{The Lowest Metallicity AGN}

In the previous section, we searched for  low
metallicity AGN by looking for objects offset from the main AGN
branch. However the diagnostic diagrams of section \ref{sec:diag}
indicate that the lowest metallicity objects could actually lie within or 
even below the star-forming branch.

To explore this possibility,  we selected from
the SDSS sample all objects with a signal-to-noise greater than 3 in
\ha, \hb\ and [\oiii]$\lambda 5007$\AA, and looked at the metallicity
sensitive line ratios such as [\sii]/\ha\, [\neiii]/\hb\ or
[\nii]/[\oii]. The
spectra, images and galaxy parameters of  all
objects with extremely low values of these ratios were examined. 

All of these objects appeared to be indistinguishable from low mass,
low metallicity starburst galaxies. They all had blue colours and high
equivalent width emission lines. None showed obvious signs of AGN
activity, such as 
the \heii\ $\lambda 4686$\AA\ emission line or [\nev]$\lambda
3426$\AA\ emission in the higher redshift objects. We found no clear
evidence for 
extremely low metallicity AGN. If such objects 
exist, they are either very rare they are indistinguishable from a 
low metallicity starburst galaxy.

\section{Discussion}\label{sec:discuss}

Low metallicity AGN appear to be very rare objects, at least
locally. The SDSS emission line galaxy sample only gives approximately
40 AGN with a mass below $10^{10}\mathrm{M}_{\odot}$ and a significant
detectable difference in their abundances relative to typical Seyfert
galaxies.  

The use of a low mass selection criterion is supported by the
observations which reveal a correlation between mass and metal
sensitive line ratios such as [\nii]/\ha. An exact determination of
a mass--metallicity relationship in AGN is hindered by the difficulty
of measuring 
with any accuracy the metallicity in the AGN hosts.

The 40 low metallicity AGN selected out of the 23000 Seyfert 2s in our
original SDSS sample  
appear to have metallicities
around half that of typical AGN. This leads to two questions: where are
the low metallicity AGN and why are there no AGN with even lower
metallicities?

We do lose some objects because we use the [\oii]$\lambda 3626$\AA\ line as a
diagnostic, which means we miss low metallicity objects below
$z\sim0.03$. Approximately 50 low mass AGN lie below this redshift,
but we estimate that only 10  would fall into the low metallicity category.
Another possibility is that low metallicity AGN are ``hidden'' by
strong star formation.                                           
As discussed in \citet{Kauffmann03b}, a
large fraction of low mass galaxies are experiencing bursts of
starformation and show strong line emission. 
However figure 6 in \citet{Kauffmann03c} demonstrates
that it is not easy to hide  a typical AGN of high metallicity.
These authors added AGN of different [\oiii] luminosities to
low mass starburst galaxies and showed that at [\oiii] luminosities
above $10^6 L_{\odot}$, all of them would have been detected.
We have repeated the same analysis for low metallicity AGN
and we find very similar results. 

It is thus clear that low metallicity AGN are rare in the local Universe,
What about at high redshift? Our emission-line diagnostics involve the 
rest-frame
optical band, and the required spectroscopy is challenging at high redshift.
Nevertheless, \citet{Shapley05} and \citet{Erb06} have recently
measured the \hb, [\oiii]$\lambda$5007, \ha, and
[\nii]$\lambda$6584 emission line fluxes in two small samples of galaxies at
$z \sim$ 1.4 and $z \sim 2.3$ respectively. We have plotted the locations of
these galaxies in Figure \ref{fig:niilowM}. Nearly all the objects lie between
the Kauffmann 
and Kewley classification lines. This region is dominated by composite
objects, where
both an AGN and star formation contribute significantly to the
emission-lines. However, the high redshift galaxies are displaced to
significantly lower [\nii]/\ha ratios than typical SDSS composite
galaxies. 

The most natural interpretation of this is that these high redshift galaxies
are indeed composite galaxies, but of lower metallicity than the typical
SDSS galaxies. Similar objects comprise only a small fraction of the SDSS
sample, but in the region where most of the high redshift galaxies are
located, bounded by the two classification curves and the main Seyfert
branch ($\log([$\oiii$]/$\hb$)\gapprox
3\log([$\nii$]/$\ha$)$), we still find about 500 SDSS galaxies. Modelling
these as composites implies that the AGN would produce 50 to 70\% of the
[\oiii]5007 flux, but only about 20\% of the \ha\ and \hb\ flux. A
definitive test of the presence of an AGN would require the detection of the
\heii $\lambda$4686 or [\nev]$\lambda$3426 lines. Unfortunately, the former
line is expected to be weak (a 20\% AGN contribution to H$\beta$ implies
\heii/\hb = 0.05), while the latter line is not redshifted into the SDSS
spectral band for the great majority of objects. 

While we can not definitively say that {\it typical} SDSS galaxies with
spectra similar to the high redshift galaxies have AGN, there are indirect
pieces of evidence that strongly suggest that this is the case. When
compared with a sample of nearby pure star-forming galaxies (all
galaxies within $\pm 0.05$ dex of [-0.55,0.10] on the \nii\ BPT
diagram), we find that 
the host galaxies of these candidate low-metallicity composites differ
in a way that is similar to what is found 
for the host galaxies of AGN \citep{Kauffmann03c,Heckman04}. 
They have on average significantly higher
stellar masses ($\sim 0.3$ dex), older stellar populations (D4000 of
1.29 compared to 1.16), redder
colors ($\sim 0.2$ greater in $g-r$) and weaker \ha\ equivalent widths
than the pure star forming
galaxies. Similar differences are found in their surface mass density,
$\sigma_{*}$, Concentration index and other galactic parameters. A 
comparison of the composites with objects just below the Kauffmann
classification curve shows similar, albeit weaker, differences.
However within the composite region 
there are some galaxies that are likely to
be pure starforming galaxies as they have strong \ha\ equivalent
widths and host properties that are similar to ordinary star-forming galaxies.

The \citet{Erb06} object at 
$\log($[\nii]/\ha$)\approx-1.0$ is the clear exception to the previous
objects. Lying above both the Kauffmann and Kewley classification
curves, it requires a much greater
contribution by an AGN to be offset from the star-forming branch due
to its low metallicity.In our SDSS AGN selected sample, we have no low
metallicity AGN or composites with 
which to compare this object. If we 
select the few SDSS objects that lie within the errors of the Erb
object, all are clear starbursting galaxies with indicators such as
the neon emission lines and \heii $\lambda 
4686$\AA\ are so weak, if present at all, as to indicate that the
contribution of an AGN to the spectrum is negligible. Thus this object
is most likely to be a low metallicity starburst.

We conclude that most of the high redshift galaxies in Figure
\ref{fig:niilowM} are most likely to be
composite objects with an AGN. Fortunately, the AGN will have only a modest
effect on the [\nii]/\ha\ ratio used by \citet{Shapley05} and
\citet{Erb06} to measure the metallicity. We estimate that the AGN contribution
will increase [\nii]/\ha\ by only about 0.1 to 0.2 dex, and hence cause
the metallicity to be overestimated by less than 0.1 dex \citep{Pettini04}.

If the \citet{Shapley05} and \citet{Erb06} galaxies are composite
AGN/starforming galaxies, this may imply 
that low metallicity AGN are more common at high redshifts. We note, however, 
that investigations
of the metallicities of QSOs at redshifts greater than 1 using both
broad and narrow emission lines all appear to
show solar to supersolar metallicities
\citep[e.g.][]{Dietrich03,Nagao06}.

The scarcity of low metallicity AGN locally is very suggestive. The
results indicate that it is unlikely for a strong AGN to be found in a
galaxy whose mass is less than $\sim 10^{10}\mathrm{M}_{\odot}$, or
whose metallicity is less than solar. It is not clear whether this
indicates that black holes only form 
in galaxies above some mass threshhold.
There is considerable evidence that high luminosity AGN are hosted
by galaxies with young stellar populations and strong
post-burst features \citep{Kauffmann03c}.
This suggests that the AGN must be preceded by strong star formation and
this may also explain the high metallicities in the NLR gas.
Further studies of AGN in low mass hosts \citep[see e.g.][]{Greene04},
and at early cosmological epochs when the mean star
formation rate and gas densities were much higher  will shed more light
on these questions.

\section*{acknowledgements}
B.G.~thanks A. Shapley for an interesting discussion about the high
redshift star-forming galaxies.

Funding for the Sloan Digital Sky Survey (SDSS) has been provided by
the Alfred P. Sloan Foundation, the Participating Institutions, the
National Aeronautics and Space Administration, the National Science
Foundation, the U.S. Department of Energy, the Japanese
Monbukagakusho, and the Max Planck Society. The SDSS Web site is
http://www.sdss.org/.

The SDSS is managed by the Astrophysical Research Consortium (ARC) for
the Participating Institutions. The Participating Institutions are the
University of Chicago, Fermilab, the Institute for Advanced Study, the
Japan Participation Group, the Johns Hopkins University, Los Alamos
National Laboratory, the Max-Planck-Institute for Astronomy (MPIA),
the Max-Planck-Institute for Astrophysics (MPA), New Mexico State
University, the University of Pittsburgh, Princeton University, the
United States Naval Observatory, and the University of Washington. 

%\small
%\singlespace


\begin{thebibliography}{}
\expandafter\ifx\csname natexlab\endcsname\relax\def\natexlab#1{#1}\fi

\bibitem[Adelman-McCarthy et al.(2006)]{DR4} 
Adelman-McCarthy, J.~K., et al.\ 2006, \apjs, 162, 38

\bibitem[Aller(1942)]{Aller42} Aller, L.~H.\ 1942, \apj,
95, 52

\bibitem[Asplund et al.(2005)]{Asplund05} Asplund, M., Grevesse, 
N., \& Sauval, A.~J.\ 2005, ASP Conf.~Ser.~336: Cosmic Abundances as 
Records of Stellar Evolution and Nucleosynthesis, 336, 25 

\bibitem[Baldwin, Phillips, \& Terlevich(1981)]{BPT81} Baldwin, J.~A., 
Phillips, M.~M., \& Terlevich, R.\ 1981, PASP, 93, 5 (BPT)

\bibitem[Baldwin et al.(1995)]{Baldwin95} Baldwin, J., Ferland, 
G., Korista, K., \& Verner, D.\ 1995, \apjl, 455, L119

\bibitem[Barth et al.(2005)]{Barth05} Barth, A.~J., Greene, 
J.~E., \& Ho, L.~C.\ 2005, \apjl, 619, L151

\bibitem[Bertoldi et al.(2003)]{Bertoldi03} Bertoldi, F., Carilli, 
C.~L., Cox, P., Fan, X., Strauss, M.~A., Beelen, A., Omont, A., \& Zylka, 
R.\ 2003, \aap, 406, L55

\bibitem[Bruzual \& Charlot(2003)]{Bruzual03} Bruzual, G., \& 
Charlot, S.\ 2003, MNRAS, 344, 1000

\bibitem[Charlot \& Longhetti(2001)]{Charlot01} Charlot, S., \& 
Longhetti, M.\ 2001, \mnras, 323, 887

\bibitem[Charlot \& Fall(2000)]{Charlot00} Charlot, S., \& Fall, 
S.~M.\ 2000, \apj, 539, 718

\bibitem[Cid Fernandes et al.(2001)]{CidFernandes01} Cid Fernandes, 
R., Heckman, T., Schmitt, H., Delgado, R.~M.~G., \& Storchi-Bergmann, T.\ 
2001, \apj, 558, 81

\bibitem[Cid Fernandes et al.(2004)]{CidFernandes04} Cid Fernandes, 
R., Gu, Q., Melnick, J., Terlevich, E., Terlevich, R., Kunth, D., Rodrigues 
Lacerda, R., \& Joguet, B.\ 2004, \mnras, 355, 273

\bibitem[Crenshaw et al.(2003)]{Crenshaw03} Crenshaw, D.~M.,
Kraemer, S.~B., \& George, I.~M.\ 2003, \araa, 41, 117

\bibitem[Davidson \& Netzer(1979)]{Davidson79} Davidson, K., \& 
Netzer, H.\ 1979, Reviews of Modern Physics, 51, 715

\bibitem[Dietrich et al.(2003)]{Dietrich03} Dietrich, M., Hamann, 
F., Shields, J.~C., Constantin, A., Heidt, J., J{\"a}ger, K., Vestergaard, 
M., \& Wagner, S.~J.\ 2003, \apj, 589, 722

\bibitem[Dopita et al.(2000)]{Dopita00}
Dopita, M.~A., Kewley, L.~J., Heisler, C.~A., \& Sutherland, R.~S.\ 2000,
\apj, 542, 224

\bibitem[Dopita \& Sutherland(2003)]{ADU03} Dopita, M.~A., \&
Sutherland, R.~S.\ 2003, Astrophysics of the diffuse universe, Berlin, New
York: Springer, 2003.~Astronomy and astrophysics library, ISBN 3540433627,

\bibitem[Elvis et al.(1994)]{Elvis94} Elvis, M., et al.\ 1994,
\apjs, 95, 1

\bibitem[Erb et al.(2006)]{Erb06} Erb, D.~K., Shapley, A.~E., Pettini,
M., Steidel, C., Reddy, N., \& Adelberger, K.\ 2006, \apj, accepted
(astro-ph/0602473) 

\bibitem[Ferrarese \& Merritt(2000)]{Ferrarese00} Ferrarese, L., \& 
Merritt, D.\ 2000, \apjl, 539, L9

\bibitem[Gallazzi et al.(2005)]{Gallazzi05} Gallazzi, A., Charlot, 
S., Brinchmann, J., White, S.~D.~M., \& Tremonti, C.~A.\ 2005, \mnras, 362, 
41

\bibitem[Gebhardt et al.(2000)]{Gebhart00} Gebhardt, K., et al.\ 
2000, \apjl, 539, L13

\bibitem[Greene \& Ho(2004)]{Greene04} Greene, J.~E., \& Ho, 
L.~C.\ 2004, \apj, 610, 722

\bibitem[Grevesse \& Sauval(1998)]{Grevesse98} Grevesse,
N.~\& Sauval, A.~J.\ 1998, Space Science Reviews, 85, 161

\bibitem[Groves et al.(2004a)]{Groves04a} Groves, B.~A., Dopita,
M.~A., \& Sutherland, R.~S.\ 2004, \apjs, 153, 9

\bibitem[Groves et al.(2004b)]{Groves04b} Groves, B.~A., Dopita,
M.~A., \& Sutherland, R.~S.\ 2004, \apjs, 153, 75

\bibitem[Groves et al.(2006)]{Groves06} Groves, B.~A., Dopita,
M.~A., \& Sutherland, R.~S.\ 2006, in prep.

\bibitem[Haehnelt et al.(1998)]{Haehnelt98} Haehnelt, M.~G., 
Natarajan, P., \& Rees, M.~J.\ 1998, \mnras, 300, 817

\bibitem[Hamann \& Ferland(1993)]{Hamann93} Hamann, F., \& 
Ferland, G.\ 1993, \apj, 418, 11

\bibitem[Hamann \& Ferland(1999)]{Hamann99} Hamann, F., \& 
Ferland, G.\ 1999, \araa, 37, 487

\bibitem[Hamann et al.(2002)]{Hamann02} Hamann, F., Korista, 
K.~T., Ferland, G.~J., Warner, C., \& Baldwin, J.\ 2002, \apj, 564,
592

\bibitem[Heckman et al.(2004)]{Heckman04} Heckman, T.~M., 
Kauffmann, G., Brinchmann, J., Charlot, S., Tremonti, C., \& White, 
S.~D.~M.\ 2004, \apj, 613, 109

\bibitem[Kauffmann \& Haehnelt(2000)]{Kauffmann00} Kauffmann, G., 
\& Haehnelt, M.\ 2000, \mnras, 311, 576

\bibitem[Kauffmann et al.(2003a)]{Kauffmann03a} Kauffmann, G., et 
al.\ 2003, MNRAS, 341, 33

\bibitem[Kauffmann et al.(2003b)]{Kauffmann03b} Kauffmann, G., et 
al.\ 2003, \mnras, 341, 54

\bibitem[Kauffmann et al.(2003c)]{Kauffmann03c} Kauffmann, G., et 
al.\ 2003, MNRAS, 346, 1055

\bibitem[Kewley \& Dopita(2002)]{Kewley02} Kewley, L.~J., \& 
Dopita, M.~A.\ 2002, \apjs, 142, 35

\bibitem[Kewley et al.(2006)]{Kewley06} Kewley, L., et 
al.\ 2006, in prep.

\bibitem[Kimura et al.(2003)]{Kimura03} Kimura, H., Mann, I., \& 
Jessberger, E.~K.\ 2003, \apj, 582, 846

\bibitem[Kennicutt, Bresolin, \& Garnett(2003)]{Kennicutt03} 
Kennicutt, R.~C., Bresolin, F., \& Garnett, D.~R.\ 2003, \apj, 591,
801

\bibitem[Komossa \& Schulz(1997)]{Komossa97} Komossa, S., \& 
Schulz, H.\ 1997, \aap, 323, 31

\bibitem[Kormendy \& Richstone(1995)]{Kormendy95} Kormendy, J., \& 
Richstone, D.\ 1995, \araa, 33, 581

\bibitem[Magorrian et al.(1998)]{Magorrian98} Magorrian, J., et 
al.\ 1998, \aj, 115, 2285

\bibitem[Matteucci \& Padovani(1993)]{Matteucci93} Matteucci, F., 
\& Padovani, P.\ 1993, \apj, 419, 485

\bibitem[Merloni et al.(2003)]{Merloni03} Merloni, A., Heinz, S., 
\& di Matteo, T.\ 2003, \mnras, 345, 1057

\bibitem[Mouhcine \& Contini(2002)]{Mouhcine02} Mouhcine, M.~\&
Contini, T.\ 2002, \aap, 389, 106

\bibitem[Nagao et al.(2006)]{Nagao06} Nagao, T., Maiolino, R., 
\& Marconi, A.\ 2006, \aap, 447, 863 

\bibitem[Nagao et al.(2002)]{Nagao02} Nagao, T., Murayama, T., 
Shioya, Y., \& Taniguchi, Y.\ 2002, \apj, 575, 721

\bibitem[Oliva et al.(1999)]{Oliva99} Oliva, E., Marconi, A., 
\& Moorwood, A.~F.~M.\ 1999, \aap, 342, 87

\bibitem[Osterbrock(1989)]{Osterbrock89}  Osterbrock, D. E., 1989,
Astrophysics of Gaseous Nebulae and Active Galactic Nuclei,
(University Science Books) 

\bibitem[Pagel et al.(1992)]{Pagel92}
Pagel, B.~E.~J., Simonson, E.~A., Terlevich, R.~J., \& Edmunds, M.~G.\
1992, \mnras, 255, 325 

\bibitem[Pettini \& Pagel(2004)]{Pettini04} Pettini, M., \& 
Pagel, B.~E.~J.\ 2004, \mnras, 348, L59 

\bibitem[Richstone et al.(1998)]{Richstone98} Richstone, D., et 
al.\ 1998, \nat, 395, A14

\bibitem[Savage \& Sembach(1996)]{Savage96} Savage, B.~D.~\&
Sembach, K.~R.\ 1996, \araa, 34, 279

\bibitem[Schmitt et al.(1999)]{Schmitt99} Schmitt, H.~R., 
Storchi-Bergmann, T., \& Fernandes, R.~C.\ 1999, \mnras, 303, 173

\bibitem[Shapley et al.(2005)]{Shapley05} Shapley, A.~E., Coil, 
A.~L., Ma, C.-P., \& Bundy, K.\ 2005, \apj, 635, 1006

\bibitem[Storchi Bergmann \& Pastoriza(1989)]{Storchi89} Storchi 
Bergmann, T., \& Pastoriza, M.~G.\ 1989, \apj, 347, 195

\bibitem[Storchi-Bergmann et al.(1998)]{Storchi98} 
Storchi-Bergmann, T., Schmitt, H.~R., Calzetti, D., \& Kinney, A.~L.\ 1998, 
\aj, 115, 909

\bibitem[Sutherland \& Dopita(1993)]{Sutherland93} Sutherland, 
R.~S., \& Dopita, M.~A.\ 1993, \apjs, 88, 253

\bibitem[Tremonti et al.(2004)]{Tremonti04} Tremonti, C.~A., et 
al.\ 2004, \apj, 613, 898

\bibitem[Vladilo(2002)]{Vladilo02} Vladilo, G.\ 2002, \apj, 569, 
295

\end{thebibliography}
\end{document}